\documentclass{aa}
\usepackage{graphicx}
\usepackage{mathptmx}   
\usepackage{amssymb}    
\usepackage{amsmath}
\usepackage{xcolor}
\usepackage[colorlinks=true,linkcolor=blue,citecolor=blue,urlcolor=blue]{hyperref}

\begin{document}
\defcitealias{Tomasetti2026}{Paper~I}

\title{Reaching the Metallicity Floor at $z\sim 10$: Lensed Star Clusters at Cosmic Dawn and Cosmic Noon}

\titlerunning{Reaching the Metallicity Floor at $z\sim10$}

 \author{Raul Jimenez\inst{1,2}\thanks{e-mail: raul.jimenez@icc.ub.edu}
          \and
          Elena Tomasetti\inst{3,4}
          \and
          Carmela Lardo\inst{3,4}
          \and
          Licia Verde\inst{1,2}}

   \institute{ICC, University of Barcelona, Mart\'i i Franqu\`es 1, 08028 Barcelona, Spain.
        \and
            ICREA, Pg. Llu\'is Companys 23, Barcelona, 08010, Spain.
        \and
            Dipartimento di Fisica e Astronomia ``Augusto Righi''--Universit\`a di Bologna, via Piero Gobetti 93/2, I-40129 Bologna, Italy.
         \and
             INAF - Osservatorio di Astrofisica e Scienza dello Spazio di Bologna, via Piero Gobetti 93/3, I-40129 Bologna, Italy.
             }

   \date{\today}

\abstract{
The origins of globular clusters (GCs) are intrinsically linked to the assembly of their host galaxies. We analyze the star-cluster populations of two strongly lensed systems that bracket
Cosmic Dawn and Cosmic Noon: the Cosmic Gems arc (GEMS) at $z=9.625$, among the first galaxies, and the Sparkler at $z=1.378$. New STARRED deconvolution photometry of GEMS provides spectral energy distributions for ten unique cluster candidates, each doubly imaged, while a homogeneous Bayesian SED-fitting analysis places both populations on a common cosmological timeline. The GEMS clusters formed at $z_{\rm form}\approx 10$--$11$ (median $z_{\rm form}\simeq10.2$), consistent with formation during the rapid assembly of halos at or above the atomic-cooling scale. Their photometry requires low metallicities: all individual clusters are consistent with $[Z/{\rm H}] \lesssim -1.2$, and the data exclude $[Z/{\rm H}]\geq-0.5$, although they cannot distinguish reliably among values below $[Z/{\rm H}] \simeq -1.5$. This conclusion is unchanged when using stellar-population models that include binary evolution---an important ingredient for the ultraviolet emission of populations this young---which yield similarly low population metallicities, $[Z/{\rm H}]=-2.2$ to $-2.7$. The formal estimate, $[Z/{\rm H}] = -2.3\pm0.3$, is therefore consistent with the metallicity floor of Milky Way GCs, although its exact value remains prior-dependent.
 The Sparkler clusters formed $\sim2.5$ Gyr later, at $z_{\rm form}\approx 2$--$3.5$ in a dwarf galaxy at Cosmic Noon, and are substantially more enriched ($[Z/{\rm H}] \approx -0.5$). Comparison with Milky Way GCs places GEMS in an exceptionally early, metal-poor regime and the Sparkler among later, more enriched populations, although neither association uniquely determines an in-situ or ex-situ origin.
Illustrative one-zone closed-box and gas-regulator calculations show that the two systems are compatible with limited pre-enrichment followed by rapid enrichment and with accretion-regulated growth. Together, they probe distinct cluster-forming environments from
Cosmic Dawn to Cosmic Noon.
}


\maketitle

\section{Introduction}
\label{sec:intro}

Globular clusters (GCs) occupy a special position in the theory of galaxy formation. They are among the densest and longest-lived stellar systems known: they host some of the oldest stellar populations in the local Universe, and their present-day ages, metallicities, abundances, masses, and spatial distributions retain information about the conditions under which their host galaxies assembled. Since the early suggestion that GCs may be relics of the first bound stellar systems \citep{PeeblesDicke1968}, and the classical use of Milky Way halo clusters to infer hierarchical assembly \citep{SearleZinn1978}, GC systems have become a central archaeological probe of galaxy formation across mass and environment \citep{Harris1996,BrodieStrader2006}. In the local Universe, the age--metallicity distribution, kinematics, and abundance patterns of GCs reveal both in-situ formation in the main progenitor and ex-situ accretion through disrupted satellites \citep{ForbesBridges2010,Leaman2013,Massari2019,Kruijssen2019}. The fossil record GCs carry is however complicated by the internal physics of dense stellar systems: most old, massive GCs host multiple stellar populations and light-element abundance variations, whose origin remains debated \citep{BastianLardo2018}. This combination of cosmological relevance and complex microphysics makes GCs unusually powerful, but also unusually demanding, tracers of early star formation.

A modern view of GC formation connects old GCs to the high-pressure, gas-rich environments that prevailed at early cosmic times. In this picture, GCs are not exceptional objects formed by an entirely separate channel, but the surviving high-mass tail of the cluster population formed during the most intense phases of galaxy growth \citep{KravtsovGnedin2005, Trenti2015GCagesLCDM,Kruijssen2015,Adamo2020}. Hydrodynamical simulations and semi-analytic models increasingly link GC formation to the same physical ingredients that regulate galaxy assembly: gas surface density, turbulent pressure, star-formation efficiency, feedback, tidal disruption, and the merger history of the host halo \citep{MuratovGnedin2010,Pfeffer2018,Kruijssen2019,Choksi2018,Reina-Campos2022,Chen2024,DeLucia2024}. These models predict that the oldest clusters should form preferentially in compact, gas-rich systems at high redshift, while later GC formation should occur in progressively more massive galaxies, whose gas reservoirs are continually replenished, enriched, and stirred by cosmological accretion and mergers. 
 The GC age--metallicity relation is therefore not expected to be a universal one-dimensional relation which can be used as a clock, but rather a projection onto the age--metallicity plane of the underlying processes governing galaxy mass growth, enrichment efficiency, inflow, outflow, and environment.

Until recently, this interpretation was tested mainly by looking backwards from nearby GC systems. This approach is powerful, but indirect: present-day clusters have survived more than ten billion years of stellar evolution, two-body relaxation, tidal shocks, orbital migration, and accretion into larger galaxies. The advent of JWST\footnote{\url{https://www.stsci.edu/jwst}}, especially when combined with strong gravitational lensing, has made it possible to observe compact star-cluster populations and candidate GC progenitors at look-back times previously inaccessible.

The Sparkler galaxy at $z=1.378$, strongly lensed by SMACS~J0723.3--7327, provided one of the first examples: compact sources around the galaxy were identified as candidate evolved GCs using JWST/NIRCam photometry, and subsequent work explored their ages, metallicities, and possible connection to the build-up of a Milky-Way-like system \citep{Mowla2022,ForbesRomanowsky2023,Adamo2023,Tomasetti2025}. At still earlier times, JWST observations of strongly lensed galaxies such as the Firefly Sparkle and the Cosmic Gems arc (hereafter GEMS) have revealed compact, massive star-forming clumps and clusters in systems observed within the first several hundred Myr of cosmic history \citep{Mowla2024,adamo_bound_2024,Messa2026,Vanzella2026}. These observations offer a direct empirical bridge between young massive clusters at high redshift and the old GC populations observed today.

In a companion paper \citep[][hereafter \citetalias{Tomasetti2026}]{Tomasetti2026} we present new deconvolution photometry of the GEMS arc obtained with the STARRED package \citep{millon_image_2024}, which yields reliable spectral energy distributions for twice as many cluster candidates as previously characterized---the five bright clusters of \citet{adamo_bound_2024}, the fainter tail sources identified spectroscopically by \citet{Messa2026}, and a newly discovered pair in the main body of the arc, for a total of ten unique clusters, each appearing in two lensed counterimages---and we derive their ages and metallicities in a cosmology-independent way, exploiting the clusters as cosmological clocks. Here we exploit the same dataset for galaxy formation. We place the GEMS and Sparkler cluster populations on a common cosmological timeline: the two systems bracket the epochs of the first galaxies and Cosmic Noon, and their newly derived metallicities allow us to test whether the two systems are consistent with the expected direction of cosmic chemical enrichment. We compare the inferred formation environments with the atomic-cooling threshold (the minimum halo mass $10^8 M_{\odot}$ at 
$z\sim10$, above which gas cools efficiently and star formation becomes sustained) and with expectations for dwarf-galaxy growth, interpret the metallicities with two limiting chemical-evolution models (a closed-box model for the first halos and an accretion-regulated gas-regulator model for Cosmic-Noon dwarfs), and compare their formation epochs and metallicities with those of the in-situ and accreted Milky Way GC populations, without assigning either high-redshift system an assembly channel.

The paper is organized as follows. Section~\ref{sec:physical_motivation} sets out the physical framework of two limiting regimes of cluster formation. Section~\ref{sec:methods} describes the data and our analysis methods. Section~\ref{sec:results} presents the results: the age--metallicity relation across the two epochs, the comparison with the atomic-cooling threshold, and the connection to the Milky Way's in-situ and ex-situ GC populations. We discuss caveats in Sect.~\ref{sec:discussion} and conclude in Sect.~\ref{sec:conclusions}. In Appendix~\ref{sec:messa_comparison} we compare our metallicities with the independent determinations available in the literature for the arc.

\section{Physical motivation: two regimes of cluster formation}
\label{sec:physical_motivation}

The metallicity of the gas from which a cluster forms is not set by cosmic
time alone: it depends on the host halo's mass, star-formation efficiency, and
accretion history, and in particular on whether the star-forming reservoir
behaves as a closed or an open system. This motivates two limiting regimes for
compact cluster formation.

In an early \emph{closed-box} regime, clusters form in rare atomic-cooling
halos ($T_{\rm vir}\simeq10^4$ K, $M_{\rm h}\sim10^8\,M_\odot$ at $z\gtrsim10$;
Sect.~\ref{sec:atomic_model}). The first clusters to form there inherit near-pristine gas (the closed-box prediction is made quantitative in Sect.~\ref{sec:closedbox}); because core-collapse enrichment then operates within a few tens of
Myr, any siblings formed even slightly later would be far more enriched.
In a later open-system regime near Cosmic Noon, the interstellar medium reaches a quasi-equilibrium metallicity set by the competition between enrichment, metal-poor inflow, and outflow: the gas-regulator picture \citep{Lilly2013}. Clusters can  therefore form at intermediate, sub-solar metallicity even several Gyr after the Big Bang.

We treat these as limiting cases in a continuous space of halo mass, accretion
rate, and star-formation timescale, not as discrete channels, and ask in
Sect.~\ref{sec:results} which regime each of our two systems resembles. The
closed-box and gas-regulator models that make these statements quantitative are
specified in Sect.~\ref{sec:methods}.
Because our sample comprises a single system in each regime, we treat the comparison throughout as a test of environmental trends and as a bracketing of the enrichment range, not as a determination of a unique formation channel for either system.

\section{Methods}
\label{sec:methods}
\subsection{Data}
\label{sec:data}
Our analysis is based on a homogeneous compilation of compact star-cluster candidates in two strongly lensed high-redshift systems: the GEMS system at \(z_{\rm obs}=9.625\) \citep{adamo_bound_2024,Messa2026} and the Sparkler system at \(z_{\rm obs}=1.378\) \citep{Mowla2022}. For the Sparkler, we adopt the stellar masses, ages, metallicities, and extinctions derived by \citet{Tomasetti2025} for its five GC candidates, with stellar masses corrected for gravitational magnification using the lensing factors of \citet{Claeyssens2023}.

The GEMS measurements are taken from \citetalias{Tomasetti2026}, which we briefly summarize here. JWST/NIRCam imaging of the arc in eight bands (F090W, F115W, F150W, F200W, F277W, F356W, F410M, F444W) was deconvolved with the two-channel pipeline STARRED, separating compact point sources from the diffuse arc emission. This procedure recovers the five bright clusters reported by \citet{adamo_bound_2024}, extracts photometry of SED quality for the four fainter tail sources previously identified by \citet{Messa2026}, and uncovers a new, sixth pair in the main body of the arc---for a total of 20 point sources, corresponding to ten unique clusters each split into two lensed counterimages by the caustic. A quality cut removes the three measurements with unreliable photometry in three or more bands (E1, H1, and G2); since at least one counterimage of every cluster survives, all ten clusters are retained.
As discussed in \citetalias{Tomasetti2026}, the sample is defined purely by detection and photometric quality, not by inferred age or metallicity, so the population statistics below are not selected on the quantities we analyze.
Physical parameters were derived in \citetalias{Tomasetti2026} with the Bayesian SED-fitting code \textsc{Bagpipes}, using the \citet{bruzual_stellar_2003} stellar-population models (hereafter BC03) with a Kroupa initial mass function, a Calzetti dust law ($A_V$ free between 0 and 4 mag); importantly, no cosmological age prior was imposed, so that the inferred ages are independent of the assumed background cosmology. The fits were performed under three star-formation-history (SFH) assumptions: a single burst, an exponentially declining model, and a delayed-$\tau$ model, all with star-formation timescales limited to 100 Myr as appropriate for GC progenitors. Throughout this paper we adopt the single-burst values as the reference case because they represent the limiting approximation of a short, nearly coeval formation episode. The alternative SFHs return systematically older ages by up to a factor $\sim2$ while the metallicities remain consistently very low \citepalias{Tomasetti2026}, and we quantify the effect of this choice on our conclusions in Sect.~\ref{sec:discussion}. The metallicity was sampled logarithmically (uniform prior in \(\log_{10} Z\)), the appropriate choice for a scale parameter constrained over several decades\ \citep[see][for a physical principle underlying such prior choices, whose importance for this measurement is quantified in Appendix~\ref{sec:messa_comparison}]{Jimenez2026prior}; we quote \([Z/{\rm H}]=\log_{10}(Z/Z_\odot)\) throughout, adopting $Z_\odot=0.02$, the solar scale of the BC03 grids. The prior extends below the lowest metallicity of the \citet{bruzual_stellar_2003} grid ($Z=0.005\,Z_\odot$, i.e., $[Z/{\rm H}]=-2.3$), where model spectra are extrapolated; the sensitivity of the population-level metallicity to the form of the prior and to its lower bound is quantified in Sect.~\ref{sec:discussion} and Appendix~\ref{sec:messa_comparison}. Stellar masses were corrected for magnification using the mean of the two lens models tabulated by \citet{Messa2026}, with the scatter between the models propagated conservatively into the mass uncertainties.
Because the mass budget and star-formation efficiencies below scale linearly with magnification, the $0.1$--$0.2$ dex scatter between the two lens models propagates directly into $\epsilon$ and  $\Sigma M_{*}$; we carry it through all baryon-budget statements and note where it dominates the error.
 As a cross-validation of the stellar templates, the entire catalogue was additionally refit with the BPASS (Binary Population and Spectral Synthesis) models \citep{Eldridge2017,StanwayEldridge2018}, which include interacting binaries and produce systematically harder and bluer ultraviolet spectra at fixed age and metallicity, under two star-formation histories: an exponentially declining model with free timescale (which converges to $\tau\simeq 65$ Myr) and one with $\tau$ fixed at 1 Myr, reproducing the configuration adopted by \citet{Messa2026}. These refits use the same photometry, quality cut, magnification treatment, logarithmic metallicity prior, and Monte Carlo combination as the reference catalogue, and are analyzed in Sect.~\ref{sec:results} and Appendix~\ref{sec:messa_comparison}, together with an injection--recovery calibration of both template families based on synthetic clusters observed through the real filter throughputs and noise properties.

For each of the ten GEMS clusters we combined the two counterimage measurements (where both survive the quality cut) by inverse-variance weighting, after verifying through Monte Carlo sampling of the asymmetric uncertainties that the counterimages are mutually consistent in both \(\log M_*\) and \([Z/{\rm H}]\). All uncertainties quoted in this paper are 16th--84th percentile ranges obtained by propagating the asymmetric measurement errors through the same Monte Carlo machinery.

An important feature of the metallicity determinations is that the individual posteriors are broad, with typical uncertainties of $^{+1.2}_{-1.0}$ dex per measurement: broadband photometry of very young stellar populations constrains metallicity only weakly. Individual values should therefore be interpreted with caution, and throughout this paper we base our conclusions on population-level statistics (the inverse-variance-weighted combination of all clusters) and on the corresponding upper bounds, which are considerably more robust than any single measurement.

\subsection{Cosmological framework and formation epochs}
\label{sec:cosmo_framework}
To compare the formation conditions of the GEMS and Sparkler clusters, which are observed at vastly different cosmic times, we place them within a common cosmological reference frame: a flat $\Lambda$CDM cosmology consistent with \citet{Planck2020}, with $H_0 = 67.4$ km s$^{-1}$ Mpc$^{-1}$, $\Omega_m = 0.315$, and $\Omega_\Lambda = 0.685$. For each cluster, the formation time follows from subtracting its measured age from the age of the Universe at its observation redshift, $t_{\rm form} = t(z_{\rm obs}) - {\rm Age}$, and is then converted into a formation redshift $z_{\rm form}$ by inverting the age--redshift relation. Because the SED ages carry no cosmological prior, this conversion is the only step at which a cosmology enters the formation redshifts. The age uncertainties are propagated through this conversion by Monte Carlo sampling, truncating each age posterior at the age of the Universe at $z_{\rm obs}$ (the SED fits themselves impose no such prior), which yields an asymmetric $z_{\rm form}$ distribution for every cluster and, by pooling them, for each population.

\subsection{Comparison with the atomic-cooling halo model}
\label{sec:atomic_model}
We place both systems on a halo mass--redshift diagram.
Our working hypothesis, the closed-box regime of Sect.~\ref{sec:physical_motivation} applied to GEMS, is that the GEMS clusters formed during the assembly of early halos near the atomic-cooling scale; the Sparkler, in contrast, should sit well above the threshold. We assess the internal consistency of this scenario by comparing the inferred formation epochs and environments of our cluster samples with the predictions of the atomic-cooling threshold model \citep{Tegmark1997,BarkanaLoeb2001,BrommYoshida2011}. This model predicts the minimum halo mass required for gas to cool efficiently via atomic hydrogen line transitions and subsequently collapse to form stars, as a function of redshift. We specifically use the theoretical predictions for a standard $\Lambda$CDM cosmology to construct a diagnostic diagram of minimum halo mass versus redshift.

 For clarity, we collect here the mass variables used in what follows. $M_{\rm atom}(z)$ is the atomic-cooling threshold mass\footnote{We use the precise $M_{\rm atom}(z_{\rm form})$ per cluster for the efficiencies $\epsilon$ and adopt the round value $M_{\rm atom}\simeq 10^8 M_{\odot}$
 as the reference halo for order-of-magnitude budget estimates.}, obtained by inverting the virial temperature--mass relation of \citet{BarkanaLoeb2001} at $T_{\rm vir}=10^{4}$ K, $M_{\rm atom}(z) \simeq 8\times10^{7}\,M_\odot\,[(1+z)/11.3]^{-3/2}$; $M_{\rm h}$ denotes a generic host halo mass, with baryonic reservoir $M_{\rm gas}=f_b M_{\rm h}$ and $f_b=\Omega_b/\Omega_m\approx0.157$; $M_*$ is the delensed stellar mass of an individual cluster (Sect.~\ref{sec:data}) and $\Sigma M_*$ the sum over the ten clusters; $M_{*,{\rm prior}}$ is the stellar mass formed in a halo before its clusters (Sect.~\ref{sec:closedbox}); $M_{\rm halo}$ is the Sparkler host halo mass estimated through the stellar-to-halo mass relation (SHMR); and $\epsilon = M_*/(f_b M_{\rm atom})$ is the per-cluster star-formation efficiency relative to the baryon budget of a threshold halo. The cluster masses are the \textsc{Bagpipes} total formed masses, including stellar remnants, which are the appropriate quantities for these baryon-budget arguments; masses in living stars are $\approx0.1$ dex lower, well within the uncertainties.
 The specific channel we adopt  for GEMS  is the merger-driven scenario  of \citet{Trenti2015GCagesLCDM}, which we summarize here before using it.
Metal-poor globular clusters form during major mergers of gas-rich halos near the atomic-cooling threshold at $z\gtrsim9$: the merger compresses the still nearly pristine gas and triggers the formation of one or a few bound clusters per event, and the statistics of these events follow from extended Press--Schechter (EPS) theory \citep{LaceyCole1993} applied to the assembly histories of halos of mass $M_{\rm atom}(z)$. The model therefore makes two predictions that we test in Sect.~\ref{sec:results}: the epoch at which the first clusters appear, given by the EPS formation-redshift distribution of threshold halos (Figs.~\ref{fig:halo_mass_zform} and \ref{fig:formation_redshift_comparison}), and their metallicity, near the floor, since the cluster-forming gas has not yet been enriched by the burst itself.
We place the two systems on this diagram in complementary ways---GEMS as a hypothesis test against the threshold, and the Sparkler as a contrast object whose host mass we estimate independently:
\begin{itemize}
    \item \textbf{GEMS population:} our primordial-burst hypothesis posits that the GEMS clusters formed within the very first halos to cross the atomic-cooling mass threshold. We therefore plot each GEMS cluster by associating its calculated formation redshift, $z_{\rm form}$, with the minimum halo mass predicted by the model at that redshift. If the hypothesis is correct, the GEMS population should trace the theoretical cooling boundary at $z_{\rm form} \gtrsim 10$, and the ratio of cluster mass to the available baryonic mass of such halos yields the local star-formation efficiency, $\epsilon = M_*/(f_b M_{\rm atom})$.
    \item \textbf{Sparkler system:} for the Sparkler GCs, which we propose formed in a more common dwarf-galaxy environment at a later time, we estimate the mass of the host halo. We anchor the host stellar mass to the direct SED-based estimates available in the literature, $M_{*}\simeq3$--$10\times10^{8}\,M_\odot$ depending on the adopted magnification, $\mu=5$--$12$ \citep{Mowla2022,Claeyssens2023}, and note that a conservative lower limit obtained internally---summing the stellar masses of the five observed GCs and scaling by a factor of ten to account for field stars---gives $M_{*,\rm total}\gtrsim5\times10^{8}\,M_\odot$, consistent with those estimates. We then employ a standard high-redshift SHMR \citep{Behroozi2013} to convert this stellar mass into a host dark matter halo mass, $M_{\rm halo}$, evaluated at the median formation redshift of the Sparkler GC population.
\end{itemize}
This comparison tests whether the inferred formation conditions for GEMS match the rare, massive, primordial halos predicted by the atomic-cooling model, and whether the Sparkler system instead occupies the region of parameter space corresponding to typical dwarf-galaxy halos assembling at Cosmic Noon, well above the contemporaneous atomic-cooling threshold.

\subsection{Modeling chemical evolution scenarios}
We interpret the two systems with  the two limiting chemical-evolution models of Sect.~\ref{sec:physical_motivation}, each tailored to the inferred formation environment.

\subsubsection{Primordial burst ``closed-box'' model for GEMS}\label{sec:closedbox}
The idea that primordial gas can condense rapidly into dense stellar systems
via Ly$\alpha$ cooling dates back to \citet{Peebles1984} and has been realized
in cosmological radiation-hydrodynamic simulations \citep[e.g.,][]{Kimm2016}.

To interpret the GEMS metallicities we adopt a ``closed-box'' description of a primordial dark matter halo of mass $\sim10^8\,M_\odot$ (the atomic-cooling threshold at
$z\gtrsim10$; Sect.~\ref{sec:atomic_model}). The available gas reservoir is
$M_{\rm gas}=f_b M_h$, with $f_b=\Omega_b/\Omega_m\approx0.157$ the universal baryon fraction, and the mean
metallicity produced by a stellar mass $M_*$ formed from it is
$Z\approx y\,M_*/M_{\rm gas}$, where $y\approx0.03$ is the net yield for
standard nucleosynthetic yields and a Kroupa IMF (adopted in all
chemical-evolution estimates below).

The prior-star-formation and enrichment estimates scale inversely with 
$y$: a factor-of-two change leaves the order-of-magnitude conclusions unchanged.

The model makes a specific prediction for the \emph{first} clusters to form in
such a halo. Before the first major burst the reservoir has been enriched only
by trace prior star formation, so these clusters should appear near the
metallicity floor of local GC systems, $[{\rm Fe/H}]\simeq-2.5$ to $-2.3$
\citep{Harris1996,Beasley2019}; this is the expectation we confront with the
GEMS data in Sect.~\ref{sec:results}.

We use the closed-box relation in two directions. First, \emph{backwards}: the
cluster metallicity measures the star formation that preceded the clusters in
the same reservoir\footnote{Because the
mixed gas mass and metal-retention efficiency are unknown, this is an
order-of-magnitude one-zone estimate rather than a model-independent upper
limit.}, $M_{*,{\rm prior}}=Z\,M_{\rm gas}/y$. Second, \emph{forwards}: the burst itself, with efficiency
$\epsilon=M_*/(f_b M_h)$, raises the remaining gas by
$\Delta Z\approx y\,\epsilon/(1-\epsilon)$ on the core-collapse timescale
(a few to a few tens of Myr). Together the two limits bracket the chemical
state of an atomic-cooling halo immediately before and after its first
cluster-forming burst, which we compare with the observed GEMS metallicities
in Sect.~\ref{sec:results}.

\subsubsection{Secular accretion ``open-system'' model for Sparkler}
For the Sparkler system, we employ an ``open-system'' gas-regulator model \citep{Lilly2013,wang2019elevationsuppressionstarformation} to describe the chemical evolution of its host dwarf galaxy. This model captures the continuous, regulated balance between gas inflow from the cosmic web (accretion), gas outflow driven by stellar feedback (winds), and gas consumption through star formation. In such a system, the interstellar medium (ISM) metallicity reaches an equilibrium value that can be approximated by the relation $Z \approx y \frac{\text{SFR}}{\text{MFR}}$, where $y$ is the effective stellar yield, SFR is the star formation rate, and MFR is the mass inflow rate of gas.

At the epoch of the Sparkler's formation ($z\sim2$--$3.5$), dwarf galaxies are characterized by high accretion rates of metal-poor gas from the intergalactic medium, often with $\mathrm{MFR} > \mathrm{SFR}$. By applying typical values for these quantities for dwarf galaxies at Cosmic Noon, we test whether this accretion-regulated equilibrium naturally produces the intermediate, sub-solar metallicities ($[Z/{\rm H}] \approx -0.5$) observed in the Sparkler GCs, which are characteristic of systems whose chemical evolution is regulated by continuous low-metallicity gas inflow.

\section{Results}
\label{sec:results}

\subsection{Properties of the two cluster populations}
\label{sec:properties}

Table~\ref{tab:summary_stats} summarizes the stellar masses, ages, and metallicities of the ten cluster candidates in the GEMS system and the five GCs in the Sparkler system; individual GEMS values are in Table~\ref{tab:gems_data}. We use `globular cluster' in the observational sense throughout; survival to $z=0$ is untested and discussed in Sect.~\ref{sec:discussion}.

\begin{table}[h!]
\centering
\caption{Median properties of the two cluster populations. Ages are measured at the respective observation redshifts of each system. $[Z/{\rm H}]$ is the logarithmic metallicity relative to solar; for GEMS we quote the inverse-variance-weighted population value with its formal uncertainty, derived from the BC03 reference catalogue and confirmed by the BPASS refits (Table~\ref{tab:crossfit}). The earlier-forming GEMS clusters are far more metal-poor than the later-forming Sparkler clusters.\label{tab:summary_stats}}
\setlength{\tabcolsep}{4pt}
\begin{tabular}{l c c c c}
\hline\hline
Population & N & Age & $[Z/{\rm H}]$ & $\log_{10} M_*$ \\
 & & (Gyr) & & ($M_\odot$) \\
\hline
GEMS ($z=9.625$) & 10 & $0.038$ & $-2.34\pm0.27$ & $6.8$ \\
Sparkler ($z=1.378$) & 5 & $1.74$ & $-0.48$ & $6.9$ \\
\hline
\end{tabular}
\end{table}

\begin{table*}
\centering
\caption{Properties of the GEMS cluster candidates (single-burst SFH; counterimages combined by inverse-variance weighting; uncertainties are 16th--84th percentile ranges from Monte Carlo propagation of the asymmetric measurement errors). Metallicity posteriors are broad, and population-level statements are more robust than individual values (see Sect.~\ref{sec:data}). The last column lists the star-formation efficiency $\epsilon = M_*/(f_b M_{\rm atom}(z_{\rm form}))$ (Sect.~\ref{sec:atomic_model}), in percent; values near or above 100\% signal that a single halo at the threshold cannot have formed the cluster (Sect.~\ref{sec:gems_acthalo}).\label{tab:gems_data}}
\begin{tabular}{lcccccc}
\hline\hline
ID & Images & $\log_{10}(M_*/M_\odot)$ & Age (Myr) & $[Z/{\rm H}]$ & $z_{\rm form}$ & $\epsilon$ (\%)\\
\hline
A & A1,A2 & $6.80^{+0.17}_{-0.18}$ & $24^{+7}_{-6}$ & $-2.33^{+0.73}_{-0.77}$ & $10.0^{+0.1}_{-0.1}$ & $48^{+23}_{-16}$\\
B & B1,B2 & $7.00^{+0.22}_{-0.22}$ & $15^{+5}_{-4}$ & $-2.39^{+0.78}_{-0.77}$ & $9.8^{+0.1}_{-0.1}$ & $75^{+47}_{-29}$\\
C & C1,C2 & $6.58^{+0.30}_{-0.29}$ & $20^{+10}_{-9}$ & $-2.34^{+0.82}_{-0.78}$ & $9.9^{+0.2}_{-0.1}$ & $29^{+28}_{-14}$\\
D & D1,D2 & $6.69^{+0.37}_{-0.38}$ & $67^{+35}_{-29}$ & $-2.38^{+0.81}_{-0.80}$ & $10.7^{+0.7}_{-0.5}$ & $42^{+59}_{-25}$\\
E & E2 & $5.54^{+0.38}_{-0.39}$ & $12^{+9}_{-9}$ & $-2.40^{+1.03}_{-0.95}$ & $9.8^{+0.1}_{-0.1}$ & $2.6^{+3.7}_{-1.5}$\\
F & F1,F2 & $5.55^{+0.17}_{-0.17}$ & $51^{+17}_{-15}$ & $-2.50^{+0.76}_{-0.75}$ & $10.4^{+0.3}_{-0.3}$ & $2.9^{+1.4}_{-0.9}$\\
G & G1 & $6.72^{+0.58}_{-0.55}$ & $41^{+35}_{-26}$ & $-2.37^{+1.16}_{-1.16}$ & $10.3^{+0.6}_{-0.4}$ & $45^{+123}_{-33}$\\
H & H2 & $7.21^{+0.54}_{-0.59}$ & $91^{+77}_{-63}$ & $-2.40^{+1.09}_{-0.98}$ & $11.2^{+1.8}_{-1.1}$ & $144^{+374}_{-106}$\\
I & I1,I2 & $7.31^{+0.31}_{-0.31}$ & $99^{+54}_{-47}$ & $-2.17^{+0.89}_{-0.88}$ & $11.3^{+1.3}_{-0.9}$ & $193^{+204}_{-100}$\\
J & J1,J2 & $7.39^{+0.25}_{-0.25}$ & $93^{+49}_{-41}$ & $-2.10^{+0.93}_{-0.86}$ & $11.2^{+1.1}_{-0.8}$ & $228^{+181}_{-101}$\\
\hline
\end{tabular}
\end{table*}

Two properties of the GEMS population underpin the analysis that follows. The clusters are extremely young at the epoch of observation, with a median age of 38 Myr, and they are very metal-poor, with an inverse-variance-weighted population metallicity of $[Z/{\rm H}] = -2.34 \pm 0.27$, i.e., $Z \simeq 0.5\%\,Z_\odot$. The individual posteriors are broad (Table~\ref{tab:gems_data}; Sect.~\ref{sec:data}), but the population-level result is robust: in every template and SFH configuration, each cluster is individually below $[Z/{\rm H}]\simeq-1.2$ to $-1.3$ at 68\% confidence, and the population value is stable at the $\leq0.4$ dex level against the choice of stellar templates
 (the BPASS refits bracket the adopted BC03 value; Table~\ref{tab:crossfit}, Appendix~\ref{sec:messa_comparison}). Injection--recovery simulations calibrate these statements against the true metallicity: the observed values are consistent with any $[Z/{\rm H}]\lesssim-1.5$, disfavor $[Z/{\rm H}]=-1$, and exclude $[Z/{\rm H}]\geq-0.5$, an exclusion that holds under either template hypothesis for the intrinsic spectra (Appendix~\ref{sec:messa_comparison}). Our per-cluster values sit 
$\sim 1.5 \sigma$ below the photometric estimates of \citet{Messa2026}; we show in Appendix~\ref{sec:messa_comparison} that this offset is driven by the metallicity prior and photometry rather than by the stellar templates.

The Sparkler clusters, observed at $z_{\rm obs}=1.378$ with much older ages (median 1.74 Gyr), are substantially more enriched, with a median metallicity of $[Z/{\rm H}] = -0.48$.

\subsection{The age--metallicity relation across two epochs}
\label{sec:amr}

To place these properties in their cosmological context, we converted the measured age of each cluster into a formation epoch as described in Sect.~\ref{sec:cosmo_framework} (the age of the Universe at observation is 0.497 Gyr for GEMS and 4.581 Gyr for the Sparkler). This reveals two distinct and widely separated epochs of cluster formation.

The GEMS clusters formed  at Cosmic Dawn, within the first $\sim 500$ Myr, with individual formation redshifts spanning $z_{\rm form} \approx 9.8$--$11.3$ (population median $z_{\rm form} = 10.2$), within $\lesssim100$ Myr of the observation epoch. The Sparkler clusters, in contrast, formed much later, during the peak epoch of cosmic star formation known as ``Cosmic Noon,'' with formation redshifts of $z_{\rm form} \approx 2$--$3.5$ (median 2.3).

\begin{figure}[h!]
\centering
\includegraphics[width=\columnwidth]{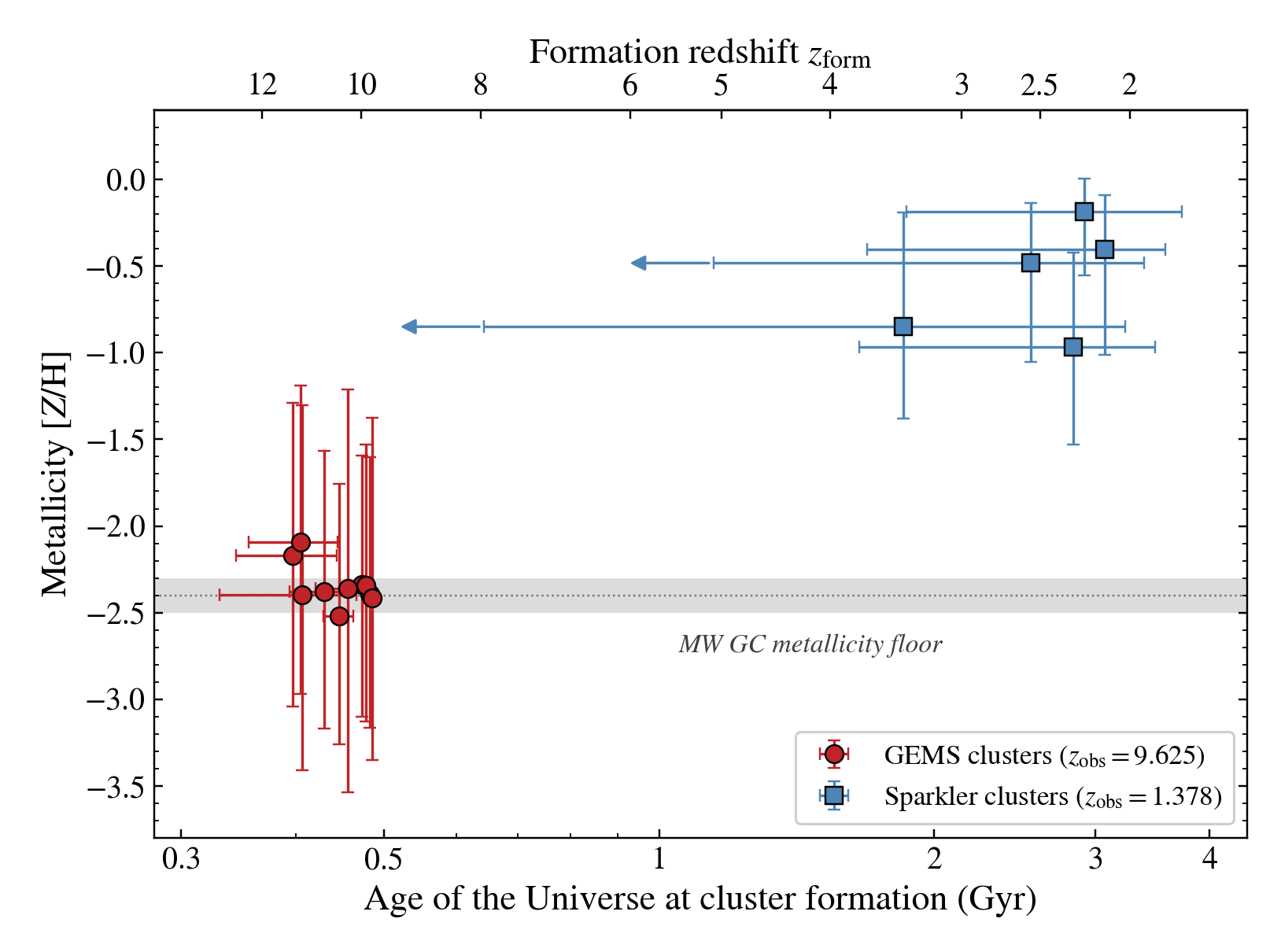}
\caption{Metallicity versus formation epoch for star clusters in the GEMS ($z_{\rm obs}=9.625$, red circles) and Sparkler ($z_{\rm obs}=1.378$, blue squares) systems; the formation epoch is expressed as the age of the Universe at cluster formation (bottom axis) and the corresponding formation redshift (top axis). GEMS points are the counterimage-combined values of Table~\ref{tab:gems_data}; Sparkler values are from \citet{Tomasetti2025}. Arrows indicate clusters whose age posteriors extend beyond the age of the Universe at $z_{\rm obs}$, leaving the early side of their formation epoch unconstrained. The grey band marks the metallicity floor of local globular-cluster systems ($[{\rm Fe/H}]\simeq-2.5$ to $-2.3$).}
\label{fig:amr}
\end{figure}

Figure~\ref{fig:amr} presents the central result of this work by combining the two measurements: metallicity against formation epoch, expressed both as the age of the Universe at cluster formation (bottom axis) and as formation redshift (top axis). Three features stand out. First, the two populations are consistent with the expected direction of cosmic chemical enrichment---the clusters that formed earliest are the most metal-poor---and their separation in metallicity is far larger than their internal scatter, so the trend is robust against the large individual uncertainties. Second, the formal GEMS population estimate lies near the metallicity floor, $[{\rm Fe/H}] \simeq -2.5$ to $-2.3$, observed for globular-cluster systems in the local Universe (Sect.~\ref{sec:physical_motivation})---a placement that is insensitive to the choice of stellar templates, with the BPASS refits straddling the floor band (Appendix~\ref{sec:messa_comparison})---though, as the injection--recovery calibration shows (Appendix~\ref{sec:messa_comparison}), the photometry cannot distinguish floor-level values from somewhat higher ones below $[Z/{\rm H}]\simeq-1.5$.
Third, clusters forming only $\sim2.5$ Gyr later carry the intermediate, sub-solar enrichment expected of star-forming dwarfs at Cosmic Noon. Together the two populations bracket the full metallicity range of the Milky Way GC system: they probe the two extremes of the enrichment conditions under which present-day GC populations were assembled, tracing the chemical growth of cluster-forming gas across the first $\sim3$ Gyr of cosmic history. As we show below, the \emph{rate} of that growth within any individual system is controlled by its formation environment rather than by cosmic time alone.

\subsection{Comparison with the atomic-cooling threshold model}

To understand the physical drivers behind these distinct formation histories, we place the two systems on the diagnostic diagram introduced in Sect.~\ref{sec:atomic_model}: in Fig.~\ref{fig:halo_mass_zform} we plot the formation redshifts of our cluster populations against their inferred host halo masses and compare their locations to the atomic-cooling threshold.  We recall from Sect.~\ref{sec:atomic_model} how the host masses are assigned: each GEMS cluster is placed at the threshold mass $M_{\rm atom}(z)$ evaluated at its formation redshift, whereas the Sparkler host halo mass is derived from the host stellar mass through the SHMR at the population's median formation redshift.

\subsubsection{GEMS: formation in the first generation of atomic-cooling halos}
\label{sec:gems_acthalo}
\begin{figure}[h!]
\centering
\includegraphics[width=\columnwidth]{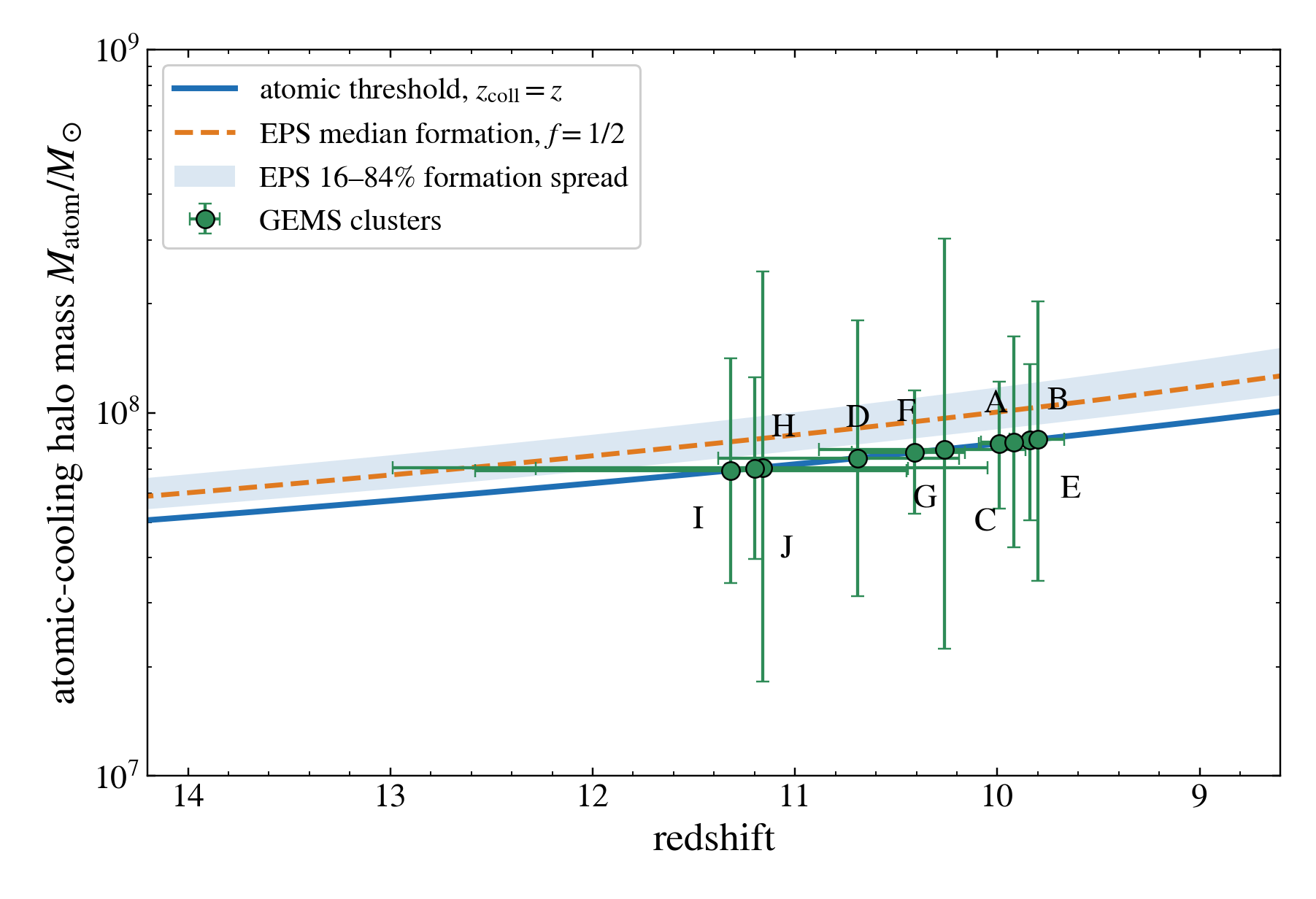}
\includegraphics[width=\columnwidth]{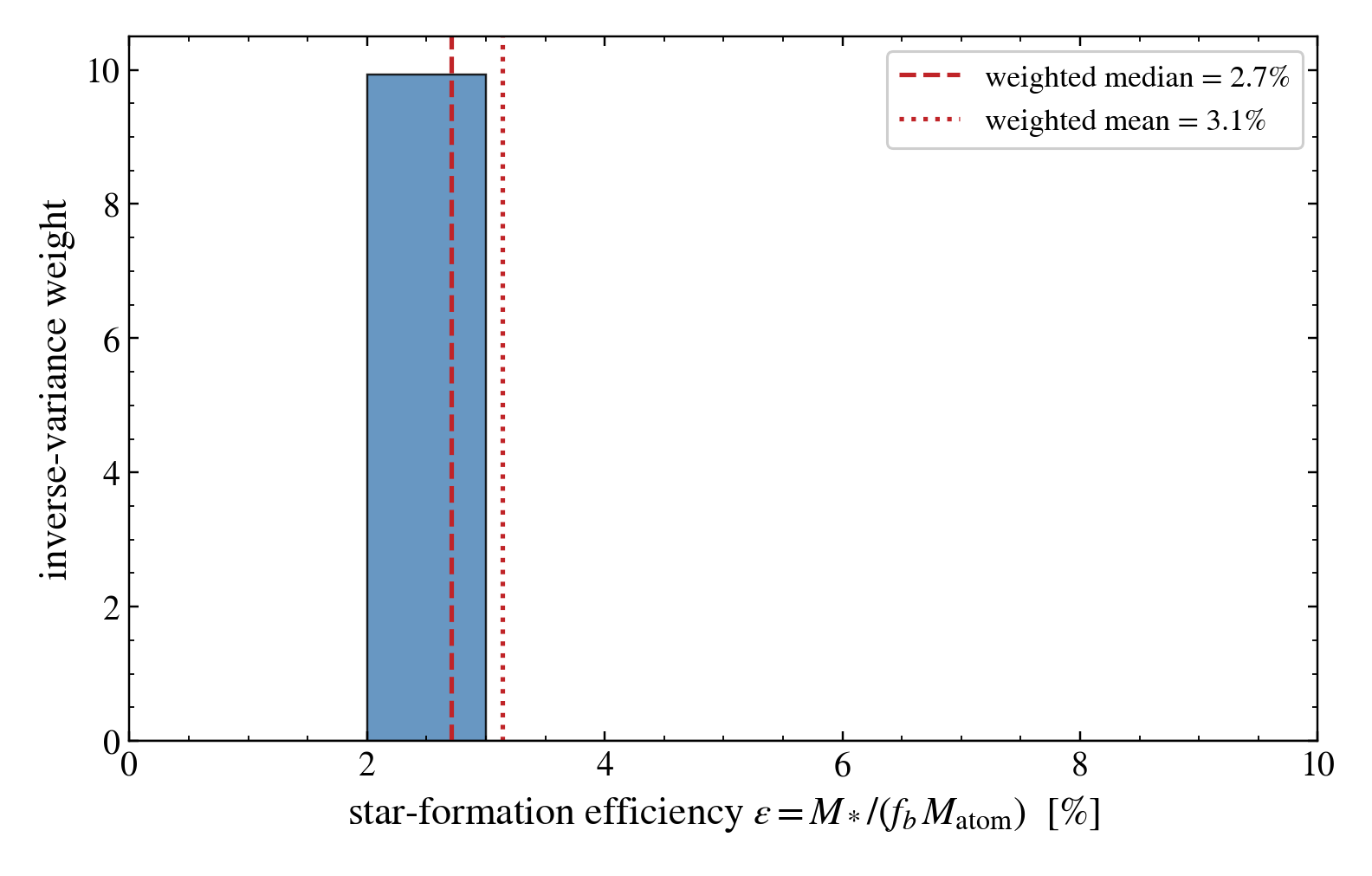}
\caption{\emph{Top:} the atomic-cooling threshold $M_{\rm atom}(z)$ ($T_{\rm vir}=10^{4}$ K; Sect.~\ref{sec:atomic_model}) (solid line), with the extended Press--Schechter median formation epoch and 16--84\% spread for halos of that mass (dashed line and shaded band). The GEMS clusters are placed on the threshold at their formation redshifts, $z_{\rm form}\approx10$--$11$ (Sect.~\ref{sec:atomic_model}); vertical bars map the stellar-mass uncertainties of Table~\ref{tab:gems_data} onto halo mass at fixed efficiency. \emph{Bottom:} inverse-variance-weighted distribution of the per-cluster efficiencies $\epsilon = M_*/(f_b M_{\rm atom})$ (Table~\ref{tab:gems_data}). The weighted median (2.7\%) and mean (3.1\%) are set by the two least massive clusters (E and F), whose small absolute uncertainties carry most of the weight; the unweighted median of the per-cluster medians is $\simeq46\%$, and the massive-cluster tail extends beyond the panel (see text).}
\label{fig:halo_mass_zform}
\end{figure}

\begin{figure}[h!]
    \centering
    \includegraphics[width=\columnwidth]{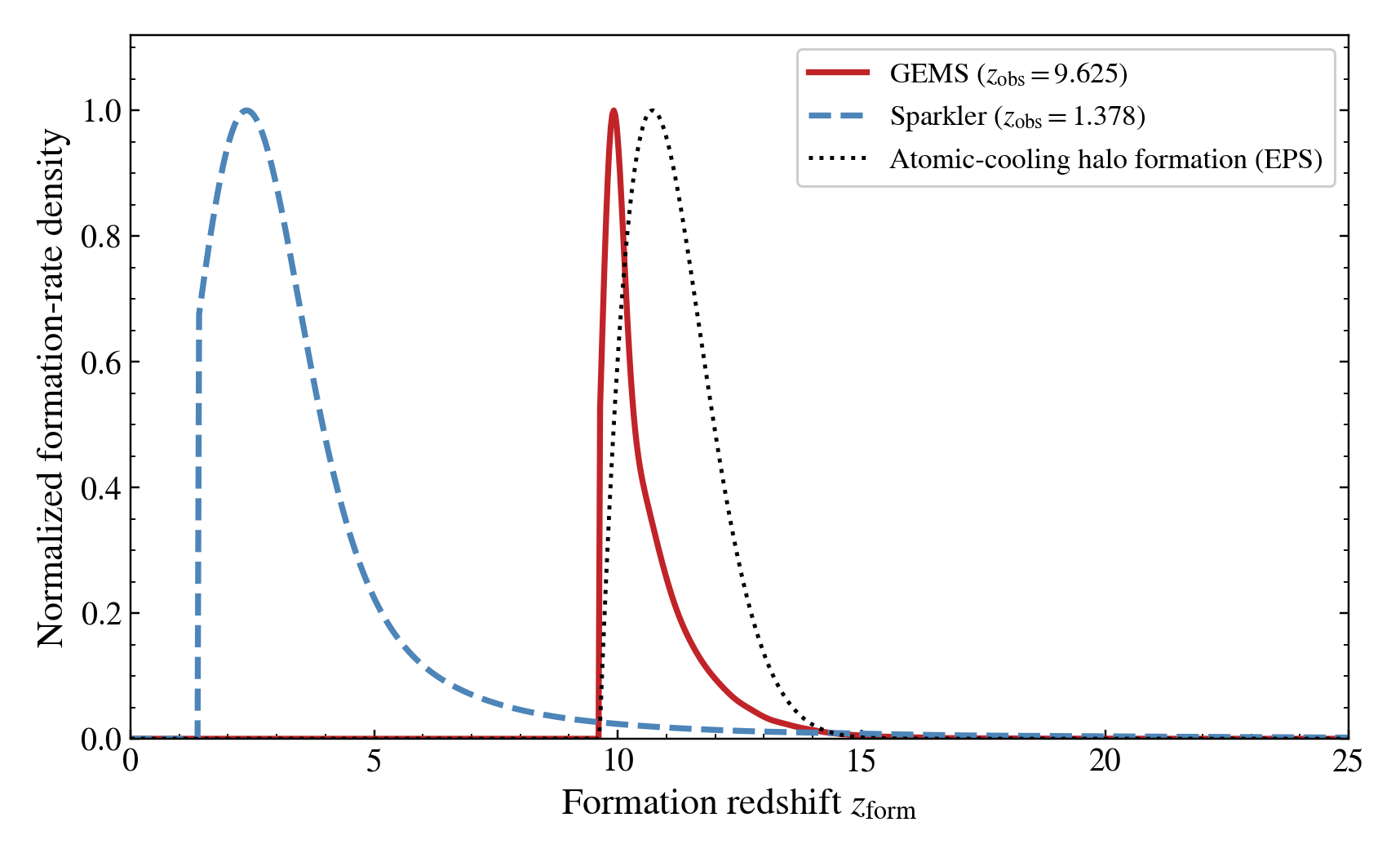}
    \caption{Formation-redshift distributions of the two cluster populations, obtained by pooling the Monte Carlo $z_{\rm form}$ posteriors of the individual clusters. The GEMS clusters ($z_{\rm obs}=9.625$, solid red) display a synchronized formation burst peaking at $z_{\rm form} \approx 10$, in agreement with the extended Press--Schechter expectation for the assembly epoch of halos at the atomic-cooling threshold (dotted black line; cf. \citealt{Trenti2015GCagesLCDM}), which peaks at $z\simeq10.7$. The GCs in the lower-redshift Sparkler system ($z_{\rm obs}=1.378$, dashed blue) formed in a later, more prolonged episode at $z_{\rm form}\approx2$--$3.5$, characteristic of hierarchical growth at Cosmic Noon.}
    \label{fig:formation_redshift_comparison}
\end{figure}

The central hypothesis is that the GEMS clusters formed within the very first dark matter halos to cross this critical mass threshold.
Because each cluster is \emph{placed} at the threshold mass $M_{\rm atom}(z_{\rm form})$ by construction (Sect.~\ref{sec:atomic_model}; Fig.~\ref{fig:halo_mass_zform}, top panel), its lying on the cooling boundary is definitional and is not evidence for the hypothesis: that placement is what is being examined. Two features of the data, neither built into the placement, bear on it. 
The stronger and more direct test is the timing: the pooled formation-redshift distribution of the population (Fig.~\ref{fig:formation_redshift_comparison}) is narrow and peaks at $z_{\rm form}\approx10$, closely matching the extended Press--Schechter expectation for the assembly epoch of halos of exactly this mass (median $z\simeq11$); this is fixed by $\Lambda$CDM, not fitted to our data. Second, and consistently, the implied efficiencies $\epsilon=M_*/(f_b M_{\rm atom})$ are physical for the low-mass clusters but exceed unity for the three most massive if each occupied a single threshold halo. 
The synchronized burst and its timing both point to the GEMS clusters being relics of the first significant wave of galaxy and star-cluster formation, born within the first rare, massive halos to ``turn on.'' The corresponding per-cluster star-formation efficiencies, $\epsilon = M_*/(f_b M_{\rm atom})$, have an inverse-variance-weighted median of $\simeq2.7\%$ (Fig.~\ref{fig:halo_mass_zform}, bottom panel), and are listed individually in Table~\ref{tab:gems_data}. The weighted median is set by the two least massive clusters (E and F): not because their photometry is superior---they are among the faintest sources---but because $\epsilon$ scales with $M_*$, so their low masses yield the smallest \emph{absolute} uncertainties on $\epsilon$ and therefore the largest inverse-variance weights. The unweighted median of the per-cluster values is instead $\simeq46\%$; the masses of Table~\ref{tab:gems_data} imply that the more massive clusters formed with substantially higher efficiencies, of several tens of percent, with the three most massive (H, I, and J) formally requiring $\epsilon\gtrsim1$ if each formed in a single halo at the threshold---an impossibility that we address below. Cluster formation in these halos was therefore locally efficient, consistent with the high formation efficiencies inferred for the bound clusters of the Cosmic Gems arc \citep{adamo_bound_2024} and with high-resolution simulations of dense cluster formation at high redshift, in which massive bound clusters form with locally high efficiencies in low-metallicity, gas-rich environments \citep[e.g.,][]{Kimm2016,Lahen2019,Lahen2020,Calura2022,Pascale2025}.  It is also consistent with the emerging picture of the arc itself as a cluster-dominated galaxy: \citet{Vanzella2026} show that its ultraviolet emission is almost entirely powered by the massive clusters, which account for a dominant fraction of the stellar mass of the host---as expected if star formation in these first halos proceeded predominantly through bound-cluster formation at the high efficiencies derived here.

We note that the ten clusters cannot have formed within a single halo at the atomic-cooling threshold: their summed delensed stellar mass,
$\Sigma M_{*} = (0.8$--$1.8) \times 10^{8}\,M_{\odot}$ (16th--84th percentiles from Monte Carlo propagation of the asymmetric mass errors of Table~\ref{tab:gems_data}, where the range is dominated by the lens-model spread rather than by photometric error), is five to eleven times the
entire baryonic reservoir of a $10^{8}\,M_{\odot}$ halo ($f_{\rm b} M_{\rm h} \simeq 1.6 \times 10^{7}\,M_{\odot}$), and the three most massive clusters individually equal or exceed it. In our
scenario the clusters instead formed in distinct atomic-cooling progenitor halos---as envisaged in the merger-driven channel of \citet{Trenti2015GCagesLCDM}---with the most massive ones requiring hosts a factor of a few above threshold or star-formation efficiencies of
several tens of percent, consistent with the upper end of the per-cluster efficiencies implied by Table~\ref{tab:gems_data}.
Summing the implied progenitor masses, $M_{{\rm h},i} = M_{*,i}/(f_{\rm b}\,\epsilon)$, the observed $z = 9.625$ host must already be a $\sim 10^{9.5}$--$10^{10}\,M_{\odot}$ halo (for
$\epsilon = 0.2$--$0.05$) that assembled its cluster-bearing progenitors within $\simeq 20$--$100$\,Myr of their formation. Such rapid merging is the defining property of the rare, high-$\sigma$ peaks in which the first atomic-cooling halos arise and, in the
\citet{Trenti2015GCagesLCDM} channel, is itself the trigger of the cluster-forming bursts. For the youngest clusters (B and E, ages $\simeq12$--$15$ Myr) the available assembly window shrinks to $\lesssim15$ Myr, comparable to a halo dynamical time at $z\simeq10$: these objects must have formed essentially during the assembly of the observed host, as expected if the mergers themselves trigger the bursts.

\subsubsection{Sparkler: formation in a more common dwarf galaxy}
The Sparkler system occupies a completely different region of this parameter space. The five observed GCs have a combined stellar mass of $\sim10^{7.7} M_\odot$. A conservative lower limit for the host galaxy's stellar mass is a factor of $\sim10$ larger, $M_{*}\gtrsim5\times10^{8} M_\odot$, and applying the SHMR of \citet{Behroozi2013} at the system's median formation redshift of $z_{\rm form} \approx 2.3$ yields a host dark matter halo mass of $M_{\rm halo}\sim10^{10.7}$--$10^{11} M_\odot$\footnote{This should be regarded as an order-of-magnitude host-halo estimate. Inferring the halo mass at the earlier GC formation epoch would require modelling the progenitor's stellar and halo mass growth.}. This is at least two orders of magnitude above the atomic-cooling threshold mass at that epoch, in a regime where halos are no longer rare density peaks but common field systems. This contrast demonstrates that the Sparkler's host was not a first-generation halo at its cooling threshold, but an ordinary dwarf galaxy that had been assembling, and enriching, for $\gtrsim2$ Gyr before these GCs formed.

\subsection{A unifying framework: two modes of cluster formation}

The distinct placement of the GEMS and Sparkler systems on the halo mass--redshift diagram and in their formation-redshift distributions (Figs.~\ref{fig:halo_mass_zform} and \ref{fig:formation_redshift_comparison}), combined with their metallicities, identifies them as fossil evidence of two distinct, epoch-dependent modes of galaxy assembly and chemical enrichment.

\subsubsection{The primordial burst mode ($z \gtrsim 10$): formation near the metallicity floor}\label{sec:burst_mode}
The GEMS clusters were born in the Universe's first deep potential wells: rare, $\sim 10^8 M_\odot$ halos at $z \gtrsim 10$, just massive enough to initiate atomic cooling (Sect.~\ref{sec:gems_acthalo}). Upon crossing this threshold, the gas in these halos collapsed and fueled a massive, centrally concentrated starburst---the event we observe as the GEMS clusters. Their metallicity constrains the point in this sequence at which they formed. Applying the closed-box relation of Sect.~\ref{sec:closedbox}, the observed population metallicity ($Z\simeq0.5\%\,Z_\odot$) corresponds to only $M_{*,{\rm prior}} \sim 5\times10^{4}\,M_\odot$ of stars---at most a few $\times10^{-3}$ of the halo gas---having formed in the reservoir before the clusters themselves.  This estimate uses the population value; since the individual posteriors are broad (Sect.~\ref{sec:data}), we also quote the conservative version: adopting the per-cluster 68\% upper bounds, $[Z/{\rm H}]\lesssim-1.2$ ($Z\lesssim6\%\,Z_\odot$), raises the limit to $M_{*,{\rm prior}}\lesssim6\times10^{5}\,M_\odot$, still only $\sim4\%$ of the halo gas reservoir.
Under the closed-box assumption this is simply the low metallicity re-expressed as a mass ($M_{*,{\rm prior}}\propto Z$ at fixed reservoir and yield), not an independent constraint: the GEMS clusters formed from gas enriched only by trace preceding star formation---a near-first stellar generation. Because it restates the metallicity rather than adding evidence, this reading is correspondingly insensitive to the residual metallicity uncertainty.

The same closed-box relation also determines the subsequent evolution. The observed clusters represent per-halo star-formation efficiencies of $\epsilon\sim3\%$ to several tens of percent,  formally reaching unity for the most massive objects if their hosts sat exactly at the threshold (Sect.~\ref{sec:gems_acthalo}; Table~\ref{tab:gems_data}); once their massive stars explode, they raise the metallicity of any retained gas to $\Delta Z \approx y\,\epsilon/(1-\epsilon) \sim 0.05$--$2\,Z_\odot$, i.e., by one to more than two orders of magnitude above the birth value, within a few tens of Myr. The low cluster metallicities are not evidence that self-enrichment had already operated: they show, on the contrary, that the clusters formed \emph{before} it. Rapid enrichment is the maximum-retention outcome of the burst; the realized enrichment depends on the fraction of ejecta retained and mixed with the surviving gas. This accounts both for the earliest surviving clusters sitting at the metallicity floor and for the gas-phase metallicities of order $0.1\,Z_\odot$ already common in galaxies observed only slightly later in cosmic time. Such rapid, burst-driven enrichment cycles---in which star formation, gas ejection, and re-accretion alternate on tens-of-Myr timescales---are a generic outcome of cosmological simulations of low-mass galaxies at these epochs \citep{McClymont2025}, and have recently been linked directly to the formation of clusters with GC-like abundance patterns \citep{McClymont2026}.

\subsubsection{The secular accretion mode ($z \approx 2$--$3.5$): ``open-system'' regulated growth}
The Sparkler GCs formed billions of years later in a much more common environment: a dwarf galaxy undergoing sustained growth during Cosmic Noon\ \citep{Mowla2022,Claeyssens2023,Adamo2023}, at the epoch when cosmological accretion and star formation peak \citep{MadauDickinson2014,Dekel2009}. Its host halo was far more massive than the minimum required for cooling at its formation epoch. The chemical evolution of such a galaxy is governed by a quasi-equilibrium process best described by ``open-system'' models. In this regime, the ISM metallicity is determined by a dynamic balance between metal production from star formation, dilution from the continuous accretion of metal-poor gas from the cosmic web, and the removal of metals via galactic winds. At $z \sim 2$--$3$, cosmological accretion rates were high, and the constant influx of pristine gas prevented the ISM from reaching solar enrichment despite the system forming several Gyr after the onset of cosmic star formation.  Quantitatively, the regulator equilibrium of Sect.~\ref{sec:methods}, $Z \approx y\,\mathrm{SFR}/\mathrm{MFR}$, with $y=0.03$ and the inflow-to-star-formation ratios $\mathrm{MFR}/\mathrm{SFR}\approx4$--$6$ typical of low-mass galaxies at these redshifts \citep[e.g.,][]{Dekel2009,Lilly2013,MaiolinoMannucci2019}, yields $Z_{\rm eq}\approx0.25$--$0.4\,Z_\odot$, i.e., $[Z/{\rm H}]\approx-0.6$ to $-0.4$, in agreement with the observed $-0.48$. The intermediate metallicities of the Sparkler GCs (Fig.~\ref{fig:amr}) are quantitatively compatible with a simplified gas-regulator equilibrium, although they do not uniquely require this regime.  The identification rests instead on the combination of an equilibrium-level metallicity, a host halo two orders of magnitude above the contemporaneous cooling threshold, and a prolonged formation epoch (Fig.~\ref{fig:formation_redshift_comparison}).

\begin{figure}[h!]
\centering
\includegraphics[width=\columnwidth]{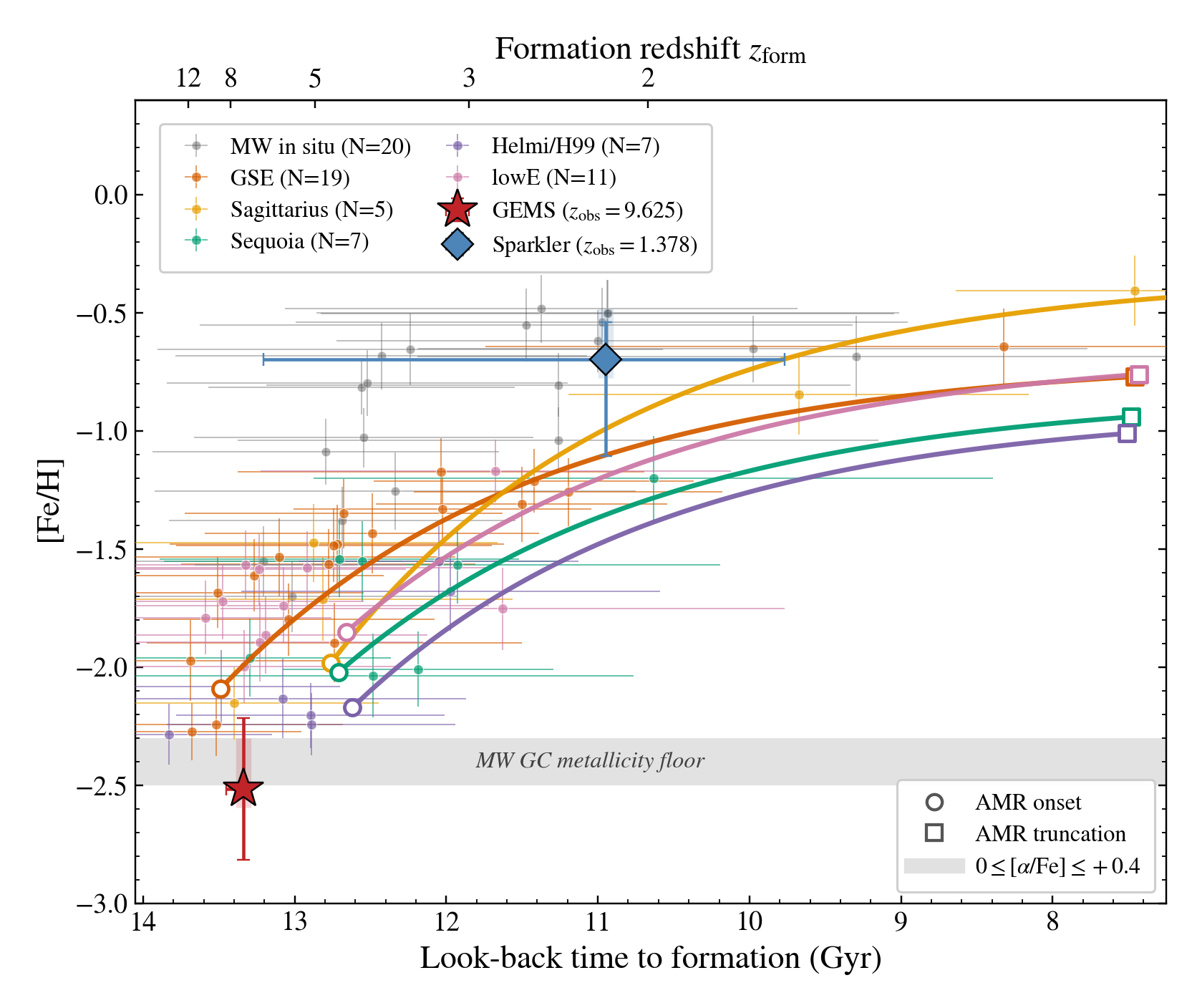}
\caption{GEMS and the Sparkler on the observed age--metallicity plane of the Milky Way GC system: 69 MW GCs with ages, [Fe/H], and progenitor classifications, and the truncated-exponential AMRs of the five accreted progenitors, from onset (circles) to truncation (squares) \citep{Lardo2026}. GEMS and Sparkler are placed at the look-back time to their formation (pooled medians with 90\% intervals; Planck18) with $[Z/{\rm H}]$ converted to [Fe/H] via the \citet{Salaris1993} relation at $[\alpha/{\rm Fe}]=+0.3$ (shaded bands: $0$ to $+0.4$); the Sparkler's metallicity bar spans the 16--84\% range of its five clusters. The grey band marks the MW GC metallicity floor. GEMS coincides with the enrichment onset of the accreted progenitors, while the Sparkler is as enriched as the Sagittarius truncation but $\sim4$ Gyr earlier, at the level of the in-situ population.}
\label{fig:mw_amr}
\end{figure}

\subsection{Comparison with the age--metallicity distributions of Milky Way GC populations}

We now compare the formation epochs and metallicities of the two systems with the observed Milky Way (MW) GC populations. This comparison identifies similarities in age--metallicity space but cannot assign an in-situ or ex-situ origin to either high-redshift system: in the Galaxy such a classification relies not only on ages and metallicities but also on chemical abundances,
orbital kinematics, and integrals of motion \citep{ForbesBridges2010,Leaman2013,
Massari2019,Kruijssen2019}. Moreover, the eventual assembly histories of GEMS and the Sparkler are unknown, and neither system is necessarily representative of the
progenitors of a MW-like galaxy. The comparison is also asymmetric in survival: the MW sample is what remains after $\sim 13$ Gyr of dynamical processing, whereas the GEMS clusters are observed at birth; the overlap in age--metallicity space therefore concerns formation conditions, not survivors.

Figure~\ref{fig:mw_amr} places the two systems on the observed age--metallicity plane of the MW GC system. The comparison sample comprises 69 MW GCs with homogeneous ages, [Fe/H], and progenitor classifications, together with the truncated-exponential age--metallicity relations (AMRs) inferred for the five
accreted progenitors---Gaia-Sausage-Enceladus (GSE), Sagittarius, Sequoia, Helmi/H99, and the low-energy group \citep{valcin25,Lardo2026}. For visual comparison, the
total metallicities of GEMS and the Sparkler are converted to illustrative $[{\rm Fe/H}]$ values with the \citet{Salaris1993} relation assuming $[\alpha/{\rm Fe}]=+0.3$; because the abundance patterns are unmeasured, the range $0\leq[\alpha/{\rm Fe}]\leq+0.4$ is shown as a systematic uncertainty. Current
GC-formation models predict substantially different high-redshift tails, with the first surviving clusters appearing at $z\simeq6$--10 in some calculations and at $z\simeq12$--14 in others \citep[see the compilation of][]{Valenzuela2025}. GEMS
therefore probes an epoch at which the models differ most strongly, although a direct model comparison must also account for selection effects and survival to $z=0$.

GEMS occupies an exceptionally early and metal-poor region of the MW GC age--metallicity plane. Its pooled formation-redshift posterior has a median $z_{\rm form}=10.2$, with 90\% of the probability at $z_{\rm form}\simeq9.7$--12.6, i.e., look-back times of $\simeq13.3$--13.5~Gyr. This median epoch falls between
the enrichment onsets reconstructed for the earliest accreted progenitors, GSE and Helmi/H99 ($\simeq13.5$ and $12.6$~Gyr). Under the low-metallicity portion of its posterior, GEMS also overlaps their inferred initial abundances, $[{\rm Fe/H}]_0\simeq-2.1$ to $-2.2$, and the most metal-poor MW GCs (e.g.\ M15, M92). The agreement in formation epoch is robust; the metallicity correspondence, however, depends on the broad, prior-sensitive GEMS posterior and on the uncertain $[{\rm Z/H}]\!\rightarrow\![{\rm Fe/H}]$ conversion, and solutions near
the conservative upper metallicity bound would lie above the AMR onset values. GEMS should therefore be regarded as an observational example of a very early, weakly enriched progenitor environment---broadly similar to some low-mass
contributors to the MW halo---rather than a system uniquely classified as ex situ.

The Sparkler occupies a later, substantially more enriched region, with a median $z_{\rm form}\simeq2.3$ (broad 90\% interval $\simeq1.6$--8.5; median look-back $\simeq10.9$~Gyr). At this epoch its metallicity lies in the enriched part of the
MW in-situ distribution and above the median AMRs of all five accreted progenitors considered here: it reaches an abundance comparable to the terminal value of Sagittarius, but some 4~Gyr earlier than in the reconstructed Sagittarius AMR, and no accreted system attains a comparable enrichment at the
same epoch. This points to more rapid and/or more efficient enrichment than in those progenitors and makes the Sparkler, in age--metallicity space, broadly analogous to the enriched MW in-situ locus; its clusters lie in the metal-rich half of the MW GC distribution and formed several Gyr after those in GEMS,
during the main epoch of galaxy and cluster formation. This analogy does not imply that the Sparkler will become the in-situ component of a present-day massive galaxy, since its host is less massive than the main MW progenitor and other galaxies may follow different enrichment and assembly histories.

In short, the two systems bracket the age--metallicity distribution of the MW GC system---GEMS at its early, metal-poor extreme and the Sparkler at a later, enriched one near the in-situ locus---as analogues of formation conditions, not
indicators of a determined assembly channel.

\section{Discussion}
\label{sec:discussion}
 It is worth stating plainly what our analysis establishes and what it does not, since our conclusions rest on three different footings. First, the robust result, independent of the metallicity prior and of the stellar templates, is that the GEMS clusters are very metal-poor: every cluster lies below
$[Z/{\rm H}]=-1.2$ at 68\% confidence, 
the population falls below both the Sparkler clusters and the arc's own nebular gas, and the injection--recovery calibration excludes a true metallicity $\gtrsim 0.3\,Z_{\odot}$.
This ordering, enrichment increasing from the first halos to Cosmic Noon, is the empirical core of the paper. Second, the specific floor-level value, $[Z/{\rm H}]\simeq -2.3$,
is prior-dependent: the photometry cannot localize the metallicity below $[Z/{\rm H}]\simeq -1.5$, 
and the point estimate follows from the (well-motivated) logarithmic prior rather than from the data alone. Third, the identification of the two systems with a primordial burst-dominated and an accretion-regulated regime is an internally consistent interpretation, not a measurement: with a single galaxy per epoch it is subject to cosmic variance, and our diagnostics test the consistency of this picture rather than excluding alternatives. The conclusions that follow rest on the first of these, are compatible with the second, and offer the third as the most natural reading.  We expand on each of these below.

The separation between the two systems in the formation epoch--metallicity plane motivates an environmental interpretation layered on top of the overall chronological enrichment sequence. The comparison is valuable not because either system is individually representative of all globular-cluster formation, but because together they isolate the dependence of cluster metallicity on formation environment: the GEMS clusters sample the onset of star formation in rare, first-generation atomic-cooling halos, while the Sparkler clusters sample the regulated, quasi-equilibrium enrichment of an ordinary Cosmic-Noon dwarf. The GEMS--Sparkler comparison thus 
 offers a concrete, if limited, empirical illustration of the expected transition from primordial burst-dominated cluster formation to accretion-regulated cluster formation across cosmic time 
(Sect.~\ref{sec:physical_motivation}).

Five caveats deserve emphasis. First, the GEMS metallicity posteriors are individually broad ($\sim1$ dex; Sect.~\ref{sec:data}); our conclusions therefore rest on the population-level combination, on the mutual consistency of the per-cluster posteriors with a single population value, and on the per-cluster upper bounds of Sect.~\ref{sec:properties}.  The likelihood is nearly flat over more than a decade in $Z$ below those bounds, so the population point value is meaningful only jointly with its stated prior: a linear-uniform prior would shift it by up to 2 dex, and the placement of the logarithmic prior's lower bound alone by up to 0.4 dex (Appendix~\ref{sec:messa_comparison}). The template systematic, in contrast, is measured rather than assumed and is subdominant ($\leq0.4$ dex; BPASS vs.\ BC03, Appendix~\ref{sec:messa_comparison})---a statement that applies to the two families tested here, BC03 and BPASS, which bracket the treatment of massive-star binarity; systematics associated with other SPS ingredients (e.g., the initial mass function, stellar rotation, or the model atmospheres at very low metallicity) remain untested. Most importantly, the injection--recovery calibration converts the posterior bounds into statements about the true metallicity---$0.3\,Z_\odot$ excluded, $0.1\,Z_\odot$ disfavored, values below $[Z/{\rm H}]\simeq-1.5$ indistinguishable from the floor---and the residual dependence on the prior's lower bound lives entirely within this calibrated allowed window. NIRSpec spectroscopy of the arc already exists, but at PRISM resolution it constrains the nebular gas rather than the stellar photospheres (Appendix~\ref{sec:messa_comparison}); deeper observations resolving stellar photospheric features would measure the birth metallicities directly and would sharpen the comparison with the local GC metallicity floor. Second, our adopted ages come from the single-burst fits; the exponentially declining and delayed-$\tau$ models explored in \citetalias{Tomasetti2026} return systematically older ages, by up to a factor of $\sim2$, while the metallicities remain low (median values of $0.2$--$1.6\%\,Z_\odot$) under all three star-formation histories; our BPASS refits confirm this directly, with median ages of 47--78 Myr and population metallicities of $-2.7$ to $-2.2$. Adopting the alternative ages would raise the GEMS formation redshifts by $\Delta z_{\rm form}\lesssim1$--$2$, pushing the population even further into the atomic-cooling regime and leaving every conclusion of this paper qualitatively unchanged. Third, the GEMS candidates are identified as globular-cluster progenitors on the basis of their present observed properties---compactness, mass, and stellar density---which say nothing about their subsequent fate: survival to $z=0$ depends on mass loss, tidal disruption, and the subsequent assembly of the host. Fourth, abundance scales for young, integrated-light metallicities need not map exactly onto the [Fe/H] scale of old Milky Way clusters; the $\alpha$-enhancement expected for rapid early star formation would shift $[Z/{\rm H}]$ relative to $[{\rm Fe/H}]$ by only a few tenths of a dex, small compared with the two-decade separation between the GEMS population and solar metallicity.

Fifth, our interpretation attributes the low GEMS metallicities to
\emph{birth} in near-pristine gas, whereas an influential alternative
attributes the local GC metallicity floor to \emph{survival}: in the framework
of \citet{Kruijssen_met_floor2019}, galaxies with $[{\rm Fe/H}]\lesssim-2.5$
are too low-mass to form clusters above the $\sim10^{5}\,M_\odot$ needed to
survive a Hubble time, so the floor would reflect a survival threshold rather
than a birth metallicity. The two pictures are not mutually exclusive, and they
are not yet separable for GEMS, whose survival to $z=0$ is untested (caveat
three); the rare sub-floor clusters known locally \citep{Wan2020,Larsen2020,%
Martin2022} may be objects that formed before their halo's first burst was
complete and only barely survived. GEMS is complementary to the survival
argument: it constrains birth metallicities at $z=9.625$ directly, before the
$>13$ Gyr of dynamical processing that the survival picture concerns.

A detailed comparison with literature metallicity estimates is presented in Appendix~\ref{sec:messa_comparison}.

\section{Conclusions}
\label{sec:conclusions}
This paper has placed two high-redshift cluster systems---GEMS at $z_{\rm obs}=9.625$ and the Sparkler at $z_{\rm obs}=1.378$---on a common cosmological timeline, using the homogeneous Bayesian determinations of their masses, ages, and metallicities from \citetalias{Tomasetti2026}. The two systems capture globular cluster formation at two widely separated epochs, and their newly derived metallicities show that the gas from which clusters form grows progressively more enriched with cosmic time.

Our analysis established that the GEMS clusters formed at $z_{\rm form}\approx10$--$11$ (population median $z_{\rm form}=10.2$). This epoch matches the parameter-free extended Press--Schechter assembly epoch of 
$\sim 10^8\,M_\odot$ halos near the atomic-cooling scale, with implied local star-formation efficiencies from a few percent to, formally, of order unity for the most massive objects. Their metallicities are very low: the population value is $[Z/{\rm H}] = -2.3\pm0.3$, with every cluster individually below $[Z/{\rm H}]\lesssim-1.2$ at 68\% confidence, a result robust to the choice of stellar templates and calibrated by injection, recovery simulations (Appendix~\ref{sec:messa_comparison}), which exclude a true metallicity $\gtrsim0.3\,Z_\odot$. This establishes that the GEMS population is very metal poor and consistent with the metallicity floor, while the available photometry cannot discriminate among metallicities below approximately $[Z/{\rm H}]\simeq-1.5$. The Sparkler clusters formed much later, at $z_{\rm form}\approx2$--$3.5$, within a more common, far more massive ($\sim10^{10.7}$--$10^{11} M_\odot$) dwarf-galaxy halo undergoing sustained assembly, and carry intermediate metallicities ($[Z/{\rm H}]\approx-0.5$).

These properties identify the two systems as fossils of two distinct, epoch-dependent modes of cluster formation:
\begin{enumerate}
    \item \textbf{A primordial burst mode ($z \gtrsim 10$):} the GEMS clusters formed at the very onset of star formation in the Universe's first deep potential wells. Their floor-level metallicity implies, through a simple closed-box argument, that only $\sim10^{4}$--$10^{5}\,M_\odot$ of stars (conservatively, $\lesssim6\times10^{5}\,M_\odot$ using the per-cluster metallicity upper bounds) preceded them in their halos---equivalently, their low metallicity expressed as a mass (Sect.~\ref{sec:burst_mode}), consistent with a near-first stellar generation. The same argument implies that the burst that formed them enriched any retained gas by one to more than two orders of magnitude within a few tens of Myr---rapid self-enrichment is the immediate consequence of this mode, not a precondition for it.
    \item \textbf{A secular accretion mode ($z \approx 2$--$3.5$):} the Sparkler clusters represent a later ``open-system'' mode of regulated growth. Their host dwarf galaxy's chemical evolution was governed by a continuous balance between star formation and the diluting inflow of metal-poor gas from the cosmic web. This accretion-regulated equilibrium naturally leads to the intermediate, sub-solar metallicities observed in the Sparkler GCs after $\gtrsim2$ Gyr of preceding enrichment.
\end{enumerate}

Although these similarities do not uniquely determine an assembly channel, in the MW age--metallicity plane, GEMS resembles the earliest, least enriched progenitor environments, whereas the Sparkler overlaps later and more enriched populations. The GEMS and Sparkler systems thus provide  direct snapshots of  the two ends of this evolution: 
from cluster formation  at Cosmic Dawn, in primordial, metal-poor atomic-cooling halos---immediately followed by violent internal enrichment---to sustained, accretion-regulated assembly at Cosmic Noon. Stellar-continuum spectroscopy of the individual clusters---deeper and at higher spectral resolution than the existing PRISM data, which constrain the post-burst nebular gas rather than the stellar photospheres---would provide a decisive test of the picture proposed here.

\begin{acknowledgements}
ET acknowledges support from the grant ASI n. 2024-10-HH.0 ``Attivit\`a scientifiche per la missione Euclid -- fase E'' and from COST Action CA21136 -- ``Addressing observational tensions in cosmology with systematics and fundamental physics (CosmoVerse)'', supported by COST (European Cooperation in Science and Technology). Funding for the work of RJ and LV was partially provided by project PID2022-141125NB-I00, and the ``Center of Excellence Maria de Maeztu 2020-2023'' award to the ICCUB (CEX2019-000918-M) funded by MCIN/AEI/10.13039/501100011033.
\end{acknowledgements}

\appendix

\section{Comparison with literature metallicity determinations of the arc}
\label{sec:messa_comparison}

Deep JWST/NIRSpec IFU spectroscopy of the Cosmic Gems arc has been
presented by \citet{Messa2026}, based on 5.8\,h of PRISM observations
($R \simeq 30$--$300$) covering $0.8$--$5.3\,\mu$m. The spectrum is
dominated by a strong Ly$\alpha$ damping break and a blue UV slope
($\beta_{\rm UV} \simeq -2.5$), with two weak emission lines
detected, H$\beta$ and [\ion{O}{iii}]$\lambda4959$, at the red edge of the
wavelength range. Three metallicity-sensitive measurements emerge from
that work, and it is essential to distinguish what each of them
traces. (i)~A \emph{gas-phase} metallicity: from the line ratio
${\rm R3} = 7.5 \pm 1.8$, \citet{Messa2026} infer
$Z_{\rm gas} \sim 10$--$20\%\,Z_{\odot}$ using the calibrations of
\citet{Nakajima2022}; we note that R3 is double-valued near its
maximum, which broadens this range, without altering the conclusions
below. The line emission is spatially concentrated near the critical curve, potentially corresponding to the two innermost sources newly identified, and may originate from gas photoionized by these compact stellar seeds ($R_{\rm{eff}}<1$ pc).
(ii)~A \emph{spectral-fit stellar} metallicity for the cluster-hosting
BCDE region, $Z_{\star} = 0.16 \pm 0.01\,Z_{\odot}$, from a
\textsc{Bagpipes} fit to the PRISM spectrum. (iii)~\emph{Per-cluster
photometric} metallicities from a re-fit of the \citet{adamo_bound_2024}
NIRCam photometry (BPASS templates, exponential SFH with
$\tau = 1$\,Myr, effectively an instantaneous burst), yielding $Z \simeq 3$--$6\%\,Z_{\odot}$, i.e.,
$[Z/{\rm H}] \simeq -1.5$ to $-1.2$, with individual uncertainties of
several tenths of a dex.

The gas-phase measurement does not trace the birth metallicity of the
clusters; on the contrary, it is a direct prediction of the
closed-box framework of Sect.~\ref{sec:closedbox}. The clusters are
young enough (ages $\lesssim100$\,Myr, and $\lesssim25$\,Myr for most
of the bright central objects; Table~\ref{tab:gems_data}) to lie
well within the core-collapse supernova enrichment window, and
the forward closed-box relation,
$\Delta Z \approx y\,\epsilon/(1-\epsilon)$, implies that the gas
surrounding them should already have been enriched to
$\sim 0.1\,Z_{\odot}$ by the time of observation. The measured
$Z_{\rm gas} \sim 10$--$20\%\,Z_{\odot}$, located within parsecs of
clusters born (in our analysis) from $\sim 0.5\%\,Z_{\odot}$ gas, is
therefore the expected signature of the immediate self-enrichment
that follows the primordial burst. The metal budget is also
consistent at the level of the whole system: the total delensed
stellar mass of the arc, $\log(M_{*}/M_{\odot}) \simeq 7.7$
\citep{Messa2026, bradley_unveiling_2025}, corresponds to
$\sim 1.5 \times 10^{6}\,M_{\odot}$ of metals for a net yield
$y \approx 0.03$, sufficient to raise $\sim 5\times10^{8}\,M_{\odot}$
of gas---comparable to the baryonic reservoir of the assembled host system---to $0.15\,Z_{\odot}$, and correspondingly less gas to higher
metallicity if mixing is incomplete  (our STARRED delensed
cluster masses sum to a larger total, $\Sigma M_{*}\simeq10^{8}\,M_\odot$,
reflecting photometry and lens-model differences with those works;
adopting the larger value only strengthens this budget). The large \ion{H}{i} column
densities measured at the cluster positions,
$\log N(\ion{H}{i})/{\rm cm}^{-2} \simeq 22.4$
\citep{Messa2026}, confirm that the
system retains abundant gas to enrich. Rather than contradicting the
low birth metallicities derived here, the nebular measurement
provides support for the forward half of the closed-box
argument: floor-metallicity cluster formation followed by rapid
enrichment of the surviving gas within a few tens of Myr.
This consistency can be made quantitative. Inverting the forward
relation, $\epsilon = \Delta Z/(y+\Delta Z)$, the measured
$Z_{\rm gas} \simeq 10$--$20\%\,Z_\odot$ corresponds to a burst
star-formation efficiency of $\epsilon \simeq 6$--$12\%$ (for
$y=0.03$ and $Z_\odot=0.02$; Sect.~\ref{sec:data})---within
the range derived independently from the cluster masses in
Sect.~\ref{sec:results}. The nebular spectroscopy and the cluster
photometry thus constrain the same event: the gas metallicity
measures the efficiency of the burst whose stars we date and
weigh.

The spectral-fit stellar metallicity of \citet{Messa2026} requires a
similar qualification. In \textsc{Bagpipes} \citep{Carnall2018} the
nebular emission model is computed at the same metallicity as the
stellar population, so in a low-resolution spectrum whose only sharp
features are H$\beta$ and [\ion{O}{iii}]$\lambda4959$ the metallicity
posterior is driven by reproducing the line fluxes---effectively by R3---while the underlying young, metal-poor stellar continuum is
nearly metallicity-independent at PRISM resolution. The quoted
precision ($\pm 0.01\,Z_{\odot}$) is characteristic of a line-ratio
constraint rather than of continuum-based stellar metallicities of
very young populations, and for the A region, where the lines are
marginally detected, the fitted metallicity indeed drops and is
described by \citet{Messa2026} as poorly constrained. We therefore
interpret measurements (i) and (ii) as two views of the same
quantity, the metallicity of the post-burst ionized gas, distinct
from the stellar birth metallicity constrained by the SED fits.

The remaining, genuinely stellar comparison is between our
STARRED-based values ($[Z/{\rm H}] \simeq -2.3$) and the per-cluster
photometric values of \citet{Messa2026}
($[Z/{\rm H}] \simeq -1.5$ to $-1.2$). Combining both error budgets,
the offset amounts to $\lesssim 1.5\sigma$ per cluster; it is
therefore not a statistically significant conflict between
measurements of individual objects, but a systematic difference whose
possible drivers are the stellar templates, the metallicity prior,
the assumed SFH, and the photometry itself.  The cross-fits
described below allow us to isolate these terms directly. Deep continuum spectroscopy sensitive
to stellar photospheric features, rather than to nebular lines,
remains the decisive future test of the birth metallicities.

To quantify the origin of this offset and the robustness of the
population-level result, we consider three consistency tests.

\emph{(i) Prior sensitivity.} An importance-reweighting estimate,
obtained by resampling the published posteriors under a linear-uniform metallicity prior truncated to the BC03 model range ($0.005$--$2.5\,Z_\odot$), yields a population value of $[Z/{\rm H}] = -0.24 \pm 0.14$, compared with $-2.3\pm0.3$ under the logarithmic prior; the same exercise reproduces, to within
$0.1$\,dex, the results of an earlier reduction of this photometry
performed with a linear prior. Likewise, moving the lower bound of
the logarithmic prior between $[Z/{\rm H}]=-2.8$ and $-4$ shifts the
population value between $-1.9$ and $-2.3$. The likelihood is
therefore nearly flat below $[Z/{\rm H}]\simeq-1$, and the population
point value is prior-dominated, as anticipated in
Sect.~\ref{sec:discussion}. Two considerations break this degeneracy
in favor of the metal-poor solution. First, the logarithmic prior is
the standard non-informative choice for a scale parameter constrained
over decades, and can be derived from first principles as
the prior consistent with the multiplicative structure of the
underlying physics \citep{Jimenez2026prior}. Second, the per-cluster values of
\citet{Messa2026}
($[Z/{\rm H}]\simeq-1.5$ to $-1.2$, from different templates, a
different SFH, and different photometry)---though themselves subject
to the biases quantified in test (iii) below---lie $\sim3\sigma$
below the linear-prior solution and, taken at face value, overlap our
logarithmic-prior posteriors within $\lesssim1.5\sigma$: whichever
way they are read, they are incompatible with a metal-rich
population.

 \emph{(ii) Template cross-fit.} We refit the
STARRED photometry with the BPASS models under the two
star-formation histories described in Sect.~\ref{sec:data}; the
results are summarized in Table~\ref{tab:crossfit} and listed per
cluster in Table~\ref{tab:bpass_percluster}. The population values,
$[Z/{\rm H}]=-2.19\pm0.30$ (free timescale) and $-2.74\pm0.21$
($\tau=1$ Myr), bracket the BC03 reference value of $-2.34\pm0.27$:
the template systematic is at most $0.4$ dex. Part of this range is
prior geometry rather than template physics: the BPASS grid extends
to $Z=5\times10^{-4}\,Z_\odot$ ($[Z/{\rm H}]\simeq-3.3$), one dex
below BC03, and truncating all posteriors at a common bound of
$[Z/{\rm H}]>-2.8$ brings the free-timescale value into exact
agreement with BC03 ($-1.9$ versus $-1.9$), while the $\tau=1$ Myr
value remains $0.4$ dex lower ($-2.3$)---a genuinely data-driven
offset, since the maximally young SFH has the bluest intrinsic
continuum and hence the strongest upper constraint on the
metallicity. The binary-enhanced
ultraviolet spectra of BPASS therefore do \emph{not} pull the
solutions to higher metallicity: on this photometry, both template
families concur that the clusters are extremely metal-poor. The
$\tau=1$ Myr configuration is effectively a single burst---the most
restrictive assumption for populations this young---and it accordingly
provides the most restrictive upper bounds: every cluster is
individually below $[Z/{\rm H}]=-1.3$ at 68\% confidence, and the
median 95\% bound is $-1.2$ (Table~\ref{tab:bpass_percluster}).

Two further conclusions follow. First, with templates and SFH now
matched to those of \citet{Messa2026},
the residual offset between our per-cluster values and theirs cannot
be attributed to the stellar library; it must reside in the
photometry (deconvolution versus aperture measurements on the direct
images) and in the metallicity sampling of the respective fits.
Their values lie at the 68\% upper bounds of our free-timescale
configuration and are therefore marginally consistent with our
posteriors. Second, the anomalously small scatter of the per-cluster
medians about the population value (0.12--0.21 dex against combined
per-cluster uncertainties of typically $\sim$0.7--0.9 dex,
$\chi^{2}=0.2$--$0.9$ for 9 degrees of freedom in all three
configurations) confirms quantitatively that the constraint is
population-wide rather than object-by-object, with the individual
medians settling coherently at the location set by the upper
constraint and the prior.

Finally, the cross-fits also confirm that the structural results of
this paper are template-independent: the BPASS masses agree with the
BC03 values object by object (worst tension $0.7\sigma$), the summed
delensed stellar mass spans $\Sigma M_{*} = (0.7$--$1.8)\times
10^{8}\,M_\odot$ across all three configurations---preserving the
mass-budget argument of Sect.~\ref{sec:results}---and the pooled
population median formation redshift spans $z_{\rm form}=10.2$--$11.3$
across the three configurations, within the star-formation-history
systematic already quoted in Sect.~\ref{sec:discussion}.

 \emph{(iii) Injection--recovery calibration.}
To calibrate what the preceding fits can and cannot measure, we
generated synthetic clusters at fixed input metallicities
$[Z/{\rm H}]_{\rm in}=-2.5$ to $-0.5$ and ages of 10, 30, and 60 Myr,
with per-source uncertainties spanning the actual signal-to-noise range of
the sample (${\rm SNR}=4$, 15, 30, and 45), and refit them with the same
machinery as the data, in all four template combinations: each family
fit with itself and each family fit with
the other.
The injected SEDs assume a single, dust-free burst, with ten noise
realizations per $(Z,\,{\rm age},\,{\rm SNR})$ node; the fits sample
the metallicity logarithmically over $Z/Z_\odot=10^{-4}$--$1.3$, the
same prior range as the fiducial analysis (Sect.~\ref{sec:data}).

The self-recovery tests establish four facts. First, in the
metal-poor regime relevant to our measurements ($[Z/{\rm
H}]_{\rm in}\leq-2$) both families recover the input to within $0.1$--$0.6$ dex at ages of 30--60 Myr (up to $\sim$0.9 dex for the youngest, 10 Myr populations), and the recovered ages are essentially unbiased at these metallicities: the ages, masses, and formation redshifts of this paper are calibrated quantities. Second, at ${\rm SNR}=4$ the
recovered metallicity is independent of the input---the faintest
sources contribute no metallicity information, as assumed throughout.
Third, a uniform (linear) metallicity prior severely degrades the
recovery for populations as young as ours: it returns $[Z/{\rm
H}]\simeq-1$ for \emph{every} input truth in BPASS, and likewise for
BC03 at 10 Myr, while retaining some sensitivity for BC03 at
30--60 Myr. Linear-prior analyses of these SEDs are therefore strongly biased toward $Z\simeq0.1\,Z_\odot$ whatever the true value. Fourth, the two
families fail differently at the metal-rich end: BC03 recovers an
input of $[Z/{\rm H}]=-0.5$ correctly ($-0.45$ to $-0.75$) at ${\rm
SNR}\geq30$ at all three ages, whereas BPASS collapses the same input
to $\simeq-2.6$ at every SNR. The upper limit therefore rests on
BC03: had the clusters been at $Z\simeq0.3\,Z_\odot$, the BC03 fits of
the bright sources would have returned that value; they returned
$-2.3$. An input of $[Z/{\rm H}]=-1$ is recovered by BC03 in an
age-dependent way (faithfully at 30 Myr and high SNR, compressed to
$\simeq-2.3$ at 10 and 60 Myr) and is therefore disfavored rather
than excluded; below $[Z/{\rm H}]\simeq-1.5$ the recovered values are
indistinguishable from those of a floor-metallicity population.

The cross-family injections close the remaining loophole: could the
low values be an artifact of fitting nature's spectra with the wrong
library? They cannot. Fitting BPASS spectra with BC03 at 30--60 Myr
\emph{compresses every input} from $-2.5$ to $-1.0$ onto $[Z/{\rm
H}]\simeq-0.65$ to $-0.95$: template mismatch in this direction biases
the result upward, toward $\simeq-1$, never downward. In the reverse
direction (BC03 spectra fit with BPASS) the recovery is approximately
faithful for metal-poor inputs at 30 Myr, compresses to $\simeq-1.2$
at 60 Myr, and collapses a metal-rich input ($-0.5$) to $\simeq-3$.
Combining the two
directions with the self-recovery results yields an exclusion that is
independent of which library describes nature: a true metallicity of
$0.3\,Z_\odot$ predicts a BC03 detection at $\simeq-0.5$ to $-0.8$
under either sky hypothesis, which is not observed. Conversely, two
distinct mechanisms---the linear prior and template mismatch---are
demonstrated to manufacture $[Z/{\rm H}]\simeq-1$ from a genuinely
metal-poor population, while no tested mechanism produces our joint
result ($-2.3$ with BC03 \emph{and} $-2.2$ to $-2.7$ with BPASS, with
mutually consistent ages) from a truth at or above $0.1\,Z_\odot$;
under mismatch, the recovered ages of the two libraries would differ
by factors of 2--5, whereas the observed
medians agree within a factor of $\lesssim2$, and within 25\% for the
SFH-matched $\tau=1$ Myr configuration. 

In summary, the calibrated statement of our measurement is: the
observed values are consistent with any true metallicity $[Z/{\rm
H}]\lesssim-1.5$, disfavor $[Z/{\rm H}]=-1$, and exclude $[Z/{\rm
H}]\geq-0.5$; within the metal-poor regime the photometry cannot
discriminate further, and the floor interpretation adopted in the
main text is the natural reading given the nebular ceiling discussed
above.

\begin{table}[h!]
\centering
\caption{Cross-fit configuration matrix for the GEMS metallicities.
All rows use the STARRED photometry, the E1/H1/G2 quality cut, and the
same counterimage combination; the first three rows use the
logarithmic metallicity prior, and the last three are linear-prior
importance-reweighting estimates of those fits, truncated to the
respective model grids. The scatter column gives the standard
deviation of the ten per-cluster medians.\label{tab:crossfit}}
\setlength{\tabcolsep}{2.5pt}\small
\begin{tabular}{l l c c c}
\hline\hline
Templates & SFH & $[Z/{\rm H}]_{\rm pop}$ & scatter & age \\
 & & & (dex) & (Myr) \\
\hline
BC03 & burst & $-2.34\pm0.27$ & 0.12 & 38 \\
BPASS & exp., $\tau$ free & $-2.19\pm0.30$ & 0.14 & 78 \\
BPASS & exp., $\tau=1$ Myr & $-2.74\pm0.21$ & 0.21 & 47 \\
\hline
BC03 & burst, lin.\ prior & $-0.24\pm0.14$ & -- & -- \\
BPASS & $\tau$ free, lin.\ prior & $-0.19$ & -- & -- \\
BPASS & $\tau=1$ Myr, lin.\ prior & $-0.6$ & -- & -- \\
\hline
\end{tabular}
\end{table}

\begin{table}[h!]
\centering
\caption{Per-cluster GEMS metallicities from the BPASS refits
(counterimages combined; 16th--84th percentile uncertainties). The
last two columns give the 68\% and 95\% upper bounds in the
$\tau=1$ Myr configuration.\label{tab:bpass_percluster}}
\setlength{\tabcolsep}{5pt}\small
\begin{tabular}{l c c c c}
\hline\hline
ID & $[Z/{\rm H}]$ & $[Z/{\rm H}]$ & \multicolumn{2}{c}{upper bounds ($\tau=1$ Myr)} \\
 & ($\tau$ free) & ($\tau=1$ Myr) & 68\% & 95\% \\
\hline
A & $-2.42^{+0.76}_{-0.74}$ & $-2.83^{+0.56}_{-0.53}$ & $-2.27$ & $-1.68$ \\
B & $-1.96^{+1.08}_{-0.94}$ & $-2.91^{+0.54}_{-0.48}$ & $-2.38$ & $-1.80$ \\
C & $-2.16^{+0.95}_{-0.84}$ & $-2.86^{+0.55}_{-0.50}$ & $-2.31$ & $-1.73$ \\
D & $-2.15^{+0.90}_{-0.81}$ & $-2.60^{+0.64}_{-0.59}$ & $-1.96$ & $-1.27$ \\
E & $-1.96^{+1.66}_{-1.29}$ & $-3.03^{+0.72}_{-0.64}$ & $-2.32$ & $-1.57$ \\
F & $-2.23^{+0.85}_{-0.81}$ & $-2.60^{+0.71}_{-0.65}$ & $-1.89$ & $-1.13$ \\
G & $-2.29^{+1.30}_{-1.11}$ & $-2.47^{+1.16}_{-0.93}$ & $-1.31$ & $-0.06$ \\
H & $-2.09^{+1.34}_{-1.12}$ & $-2.54^{+1.09}_{-0.97}$ & $-1.45$ & $-0.30$ \\
I & $-2.12^{+0.95}_{-0.87}$ & $-2.40^{+0.80}_{-0.73}$ & $-1.60$ & $-0.77$ \\
J & $-2.21^{+0.91}_{-0.85}$ & $-2.62^{+0.72}_{-0.64}$ & $-1.90$ & $-1.14$ \\
\hline
\end{tabular}
\end{table}


\begin{thebibliography}{65}
\expandafter\ifx\csname natexlab\endcsname\relax\def\natexlab#1{#1}\fi

\bibitem[{Adamo {et~al.}(2024)Adamo, Bradley, Vanzella, Claeyssens, Welch,
  Diego, Mahler, Oguri, Sharon, {Abdurro’uf}, Hsiao, Xu, Messa, Lassen,
  Zackrisson, Brammer, Coe, Kokorev, Ricotti, Zitrin, Fujimoto, Inoue,
  Resseguier, Rigby, Jiménez-Teja, Windhorst, Hashimoto, \&
  Tamura}]{adamo_bound_2024}
Adamo, A., Bradley, L.~D., Vanzella, E., {et~al.} 2024, Nature, 632, 513

\bibitem[{Adamo {et~al.}(2023)Adamo, Usher, Pfeffer, \& Claeyssens}]{Adamo2023}
Adamo, A., Usher, C., Pfeffer, J., \& Claeyssens, A. 2023, Monthly Notices of
  the Royal Astronomical Society: Letters, 525, L6

\bibitem[{Adamo {et~al.}(2020)Adamo, Zeidler, Kruijssen, Chevance, Gouliermis,
  Hollyhead, Krumholz, Larsen, Reina-Campos, Ryon, {et~al.}}]{Adamo2020}
Adamo, A., Zeidler, P., Kruijssen, J. M.~D., {et~al.} 2020, Space Science
  Reviews, 216, 69

\bibitem[{Barkana \& Loeb(2001)}]{BarkanaLoeb2001}
Barkana, R. \& Loeb, A. 2001, Physics Reports, 349, 125

\bibitem[{Bastian \& Lardo(2018)}]{BastianLardo2018}
Bastian, N. \& Lardo, C. 2018, Annual Review of Astronomy and Astrophysics, 56,
  83

\bibitem[{{Beasley} {et~al.}(2019){Beasley}, {Leaman}, {Gallart}, {Larsen},
  {Battaglia}, {Monelli}, \& {Pedreros}}]{Beasley2019}
{Beasley}, M.~A., {Leaman}, R., {Gallart}, C., {et~al.} 2019, \mnras, 487, 1986

\bibitem[{{Behroozi} {et~al.}(2013){Behroozi}, {Wechsler}, \&
  {Conroy}}]{Behroozi2013}
{Behroozi}, P.~S., {Wechsler}, R.~H., \& {Conroy}, C. 2013, ApJ, 770, 57

\bibitem[{Bradley {et~al.}(2025)Bradley, Adamo, Vanzella, Sharon, Brammer, Coe,
  Diego, Kokorev, Mahler, Oguri, {Abdurro'uf}, Bhatawdekar, Christensen,
  Fujimoto, Hashimoto, Hsiao, Inoue, Jiménez-Teja, Messa, Norman, Ricotti,
  Tamura, Windhorst, Xu, \& Zitrin}]{bradley_unveiling_2025}
Bradley, L.~D., Adamo, A., Vanzella, E., {et~al.} 2025, The Astrophysical
  Journal, 991, 32

\bibitem[{Brodie \& Strader(2006)}]{BrodieStrader2006}
Brodie, J.~P. \& Strader, J. 2006, Annual Review of Astronomy and Astrophysics,
  44, 193

\bibitem[{Bromm \& Yoshida(2011)}]{BrommYoshida2011}
Bromm, V. \& Yoshida, N. 2011, Annual Review of Astronomy and Astrophysics, 49,
  373

\bibitem[{Bruzual \& Charlot(2003)}]{bruzual_stellar_2003}
Bruzual, G. \& Charlot, S. 2003, MNRAS, 344, 1000

\bibitem[{{Calura} {et~al.}(2022){Calura}, {Lupi}, {Rosdahl}, {Vanzella},
  {Meneghetti}, {Rosati}, {Vesperini}, {Lacchin}, {Pascale}, \&
  {Gilli}}]{Calura2022}
{Calura}, F., {Lupi}, A., {Rosdahl}, J., {et~al.} 2022, \mnras, 516, 5914

\bibitem[{{Carnall} {et~al.}(2018){Carnall}, {McLure}, {Dunlop}, \&
  {Dav{\'e}}}]{Carnall2018}
{Carnall}, A.~C., {McLure}, R.~J., {Dunlop}, J.~S., \& {Dav{\'e}}, R. 2018,
  \mnras, 480, 4379

\bibitem[{{Chen} \& {Gnedin}(2024)}]{Chen2024}
{Chen}, Y. \& {Gnedin}, O.~Y. 2024, The Open Journal of Astrophysics, 7, 23

\bibitem[{{Choksi} {et~al.}(2018){Choksi}, {Gnedin}, \& {Li}}]{Choksi2018}
{Choksi}, N., {Gnedin}, O.~Y., \& {Li}, H. 2018, \mnras, 480, 2343

\bibitem[{{Claeyssens} {et~al.}(2023){Claeyssens}, {Adamo}, {Richard},
  {Mahler}, {Messa}, \& {Dessauges-Zavadsky}}]{Claeyssens2023}
{Claeyssens}, A., {Adamo}, A., {Richard}, J., {et~al.} 2023, MNRAS, 520, 2180

\bibitem[{{De Lucia} {et~al.}(2024){De Lucia}, {Kruijssen}, {Trujillo-Gomez},
  {Hirschmann}, \& {Xie}}]{DeLucia2024}
{De Lucia}, G., {Kruijssen}, J.~M.~D., {Trujillo-Gomez}, S., {Hirschmann}, M.,
  \& {Xie}, L. 2024, \mnras, 530, 2760

\bibitem[{Dekel {et~al.}(2009)Dekel, Birnboim, Engel, Freundlich, Goerdt,
  Mumcuoglu, Neistein, Pichon, Teyssier, \& Zinger}]{Dekel2009}
Dekel, A., Birnboim, Y., Engel, G., {et~al.} 2009, Nature, 457, 451

\bibitem[{{Eldridge} {et~al.}(2017){Eldridge}, {Stanway}, {Xiao}, {McClelland},
  {Taylor}, {Ng}, {Greis}, \& {Bray}}]{Eldridge2017}
{Eldridge}, J.~J., {Stanway}, E.~R., {Xiao}, L., {et~al.} 2017, Publications of
  the Astronomical Society of Australia, 34, e058

\bibitem[{Forbes \& Bridges(2010)}]{ForbesBridges2010}
Forbes, D.~A. \& Bridges, T. 2010, Monthly Notices of the Royal Astronomical
  Society, 404, 1203

\bibitem[{Forbes \& Romanowsky(2023)}]{ForbesRomanowsky2023}
Forbes, D.~A. \& Romanowsky, A.~J. 2023, Monthly Notices of the Royal
  Astronomical Society: Letters, 520, L58

\bibitem[{Harris(1996)}]{Harris1996}
Harris, W.~E. 1996, The Astronomical Journal, 112, 1487

\bibitem[{{Jimenez} {et~al.}(2026){Jimenez}, {Pe{\~n}a Garay}, {Simpson}, \&
  {Verde}}]{Jimenez2026prior}
{Jimenez}, R., {Pe{\~n}a Garay}, C., {Simpson}, F., \& {Verde}, L. 2026,
  submitted to JCAP, arXiv:2606.18491

\bibitem[{{Kimm} {et~al.}(2016){Kimm}, {Cen}, {Rosdahl}, \& {Yi}}]{Kimm2016}
{Kimm}, T., {Cen}, R., {Rosdahl}, J., \& {Yi}, S.~K. 2016, \apj, 823, 52

\bibitem[{Kravtsov \& Gnedin(2005)}]{KravtsovGnedin2005}
Kravtsov, A.~V. \& Gnedin, O.~Y. 2005, The Astrophysical Journal, 623, 650

\bibitem[{Kruijssen(2015)}]{Kruijssen2015}
Kruijssen, J. M.~D. 2015, Monthly Notices of the Royal Astronomical Society,
  454, 1658

\bibitem[{{Kruijssen}(2019)}]{Kruijssen_met_floor2019}
{Kruijssen}, J.~M.~D. 2019, \mnras, 486, L20

\bibitem[{Kruijssen {et~al.}(2019)Kruijssen, Pfeffer, Reina-Campos, Crain, \&
  Bastian}]{Kruijssen2019}
Kruijssen, J. M.~D., Pfeffer, J.~L., Reina-Campos, M., Crain, R.~A., \&
  Bastian, N. 2019, Monthly Notices of the Royal Astronomical Society, 486,
  3180

\bibitem[{{Lacey} \& {Cole}(1993)}]{LaceyCole1993}
{Lacey}, C. \& {Cole}, S. 1993, \mnras, 262, 627

\bibitem[{{Lah{\'e}n} {et~al.}(2019){Lah{\'e}n}, {Naab}, {Johansson},
  {Elmegreen}, {Hu}, \& {Walch}}]{Lahen2019}
{Lah{\'e}n}, N., {Naab}, T., {Johansson}, P.~H., {et~al.} 2019, \apjl, 879, L18

\bibitem[{{Lah{\'e}n} {et~al.}(2020){Lah{\'e}n}, {Naab}, {Johansson},
  {Elmegreen}, {Hu}, {Walch}, {Steinwandel}, \& {Moster}}]{Lahen2020}
{Lah{\'e}n}, N., {Naab}, T., {Johansson}, P.~H., {et~al.} 2020, \apj, 891, 2

\bibitem[{{Lardo} {et~al.}(2026){Lardo}, {Valcin}, \& {Jimenez}}]{Lardo2026}
{Lardo}, C., {Valcin}, D., \& {Jimenez}, R. 2026, \aap, 709, A280

\bibitem[{{Larsen} {et~al.}(2020){Larsen}, {Romanowsky}, {Brodie}, \&
  {Wasserman}}]{Larsen2020}
{Larsen}, S.~S., {Romanowsky}, A.~J., {Brodie}, J.~P., \& {Wasserman}, A. 2020,
  Science, 370, 970

\bibitem[{Leaman {et~al.}(2013)Leaman, VandenBerg, \& Mendel}]{Leaman2013}
Leaman, R., VandenBerg, D.~A., \& Mendel, J.~T. 2013, Monthly Notices of the
  Royal Astronomical Society, 436, 122

\bibitem[{Lilly {et~al.}(2013)Lilly, Carollo, Pipino, Renzini, \&
  Peng}]{Lilly2013}
Lilly, S.~J., Carollo, C.~M., Pipino, A., Renzini, A., \& Peng, Y.-j. 2013, The
  Astrophysical Journal, 772, 119

\bibitem[{Madau \& Dickinson(2014)}]{MadauDickinson2014}
Madau, P. \& Dickinson, M. 2014, Annual Review of Astronomy and Astrophysics,
  52, 415

\bibitem[{Maiolino \& Mannucci(2019)}]{MaiolinoMannucci2019}
Maiolino, R. \& Mannucci, F. 2019, Astronomy and Astrophysics Review, 27, 3

\bibitem[{{Martin} {et~al.}(2022){Martin}, {Venn}, {Aguado}, {Starkenburg},
  {Gonz{\'a}lez Hern{\'a}ndez}, {Ibata}, {Bonifacio}, {Caffau}, {Sestito},
  {Arentsen}, {Allende Prieto}, {Carlberg}, {Fabbro}, {Fouesneau}, {Hill},
  {Jablonka}, {Kordopatis}, {Lardo}, {Malhan}, {Mashonkina}, {McConnachie},
  {Navarro}, {S{\'a}nchez-Janssen}, {Thomas}, {Yuan}, \&
  {Mucciarelli}}]{Martin2022}
{Martin}, N.~F., {Venn}, K.~A., {Aguado}, D.~S., {et~al.} 2022, \nat, 601, 45

\bibitem[{Massari {et~al.}(2019)Massari, Koppelman, \& Helmi}]{Massari2019}
Massari, D., Koppelman, H.~H., \& Helmi, A. 2019, Astronomy \& Astrophysics,
  630, L4

\bibitem[{{McClymont} {et~al.}(2026){McClymont}, {Belokurov}, {Tacchella},
  {Kannan}, {Smith}, {Puchwein}, {Garaldi}, {Vogelsberger}, {Monty}, {Borrow},
  {Keating}, {Shen}, {Wang}, {Zier}, {Cho}, {Isobe}, {Ji}, {Maiolino}, \&
  {Pruto}}]{McClymont2026}
{McClymont}, W., {Belokurov}, V., {Tacchella}, S., {et~al.} 2026, arXiv
  e-prints, arXiv:2607.05509

\bibitem[{{McClymont} {et~al.}(2025){McClymont}, {Tacchella}, {Smith},
  {Kannan}, {Puchwein}, {Borrow}, {Garaldi}, {Keating}, {Vogelsberger}, {Zier},
  {Shen}, {Popovic}, \& {Simmonds}}]{McClymont2025}
{McClymont}, W., {Tacchella}, S., {Smith}, A., {et~al.} 2025, \mnras, 544, 513

\bibitem[{Messa {et~al.}(2026)Messa, Vanzella, Loiacono, Adamo, Oguri, Sharon,
  Bradley, Christensen, Claeyssens, Richard, {et~al.}}]{Messa2026}
Messa, M., Vanzella, E., Loiacono, F., {et~al.} 2026, Astronomy \&
  Astrophysics, 705, A173

\bibitem[{Millon {et~al.}(2024)Millon, Michalewicz, Dux, Courbin, \&
  Marshall}]{millon_image_2024}
Millon, M., Michalewicz, K., Dux, F., Courbin, F., \& Marshall, P.~J. 2024, AJ,
  168, 55

\bibitem[{Mowla {et~al.}(2024)Mowla, Iyer, Asada, Desprez, Tan, Martis,
  Sarrouh, Strait, Abraham, Brada{\v c}, {et~al.}}]{Mowla2024}
Mowla, L., Iyer, K., Asada, Y., {et~al.} 2024, Nature, 636, 332

\bibitem[{Mowla {et~al.}(2022)Mowla, Iyer, Desprez, Estrada-Carpenter, Martis,
  Noirot, Sarrouh, Strait, Asada, Abraham, {et~al.}}]{Mowla2022}
Mowla, L., Iyer, K.~G., Desprez, G., {et~al.} 2022, The Astrophysical Journal
  Letters, 937, L35

\bibitem[{Muratov \& Gnedin(2010)}]{MuratovGnedin2010}
Muratov, A.~L. \& Gnedin, O.~Y. 2010, The Astrophysical Journal, 718, 1266

\bibitem[{{Nakajima} {et~al.}(2022){Nakajima}, {Ouchi}, {Xu}, {Rauch},
  {Harikane}, {Nishigaki}, {Isobe}, {Kusakabe}, {Nagao}, {Ono}, {Onodera},
  {Sugahara}, {Kim}, {Komiyama}, {Lee}, \& {Zahedy}}]{Nakajima2022}
{Nakajima}, K., {Ouchi}, M., {Xu}, Y., {et~al.} 2022, \apjs, 262, 3

\bibitem[{{Pascale} {et~al.}(2025){Pascale}, {Calura}, {Vesperini}, {Rosdahl},
  {Nipoti}, {Giunchi}, {Lacchin}, {Lupi}, {Messa}, {Meneghetti}, {Ragagnin},
  {Vanzella}, \& {Zanella}}]{Pascale2025}
{Pascale}, R., {Calura}, F., {Vesperini}, E., {et~al.} 2025, \aap, 699, A31

\bibitem[{{Peebles}(1984)}]{Peebles1984}
{Peebles}, P.~J.~E. 1984, \apj, 277, 470

\bibitem[{Peebles \& Dicke(1968)}]{PeeblesDicke1968}
Peebles, P. J.~E. \& Dicke, R.~H. 1968, The Astrophysical Journal, 154, 891

\bibitem[{Pfeffer {et~al.}(2018)Pfeffer, Kruijssen, Crain, \&
  Bastian}]{Pfeffer2018}
Pfeffer, J., Kruijssen, J. M.~D., Crain, R.~A., \& Bastian, N. 2018, Monthly
  Notices of the Royal Astronomical Society, 475, 4309

\bibitem[{{Planck Collaboration} {et~al.}(2020){Planck Collaboration}, Aghanim,
  Akrami, Ashdown, Aumont, Baccigalupi, {et~al.}}]{Planck2020}
{Planck Collaboration}, Aghanim, N., Akrami, Y., {et~al.} 2020, Astronomy \&
  Astrophysics, 641, A6

\bibitem[{{Reina-Campos} {et~al.}(2022){Reina-Campos}, {Keller}, {Kruijssen},
  {Gensior}, {Trujillo-Gomez}, {Jeffreson}, {Pfeffer}, \&
  {Sills}}]{Reina-Campos2022}
{Reina-Campos}, M., {Keller}, B.~W., {Kruijssen}, J.~M.~D., {et~al.} 2022,
  \mnras, 517, 3144

\bibitem[{{Salaris} {et~al.}(1993){Salaris}, {Chieffi}, \&
  {Straniero}}]{Salaris1993}
{Salaris}, M., {Chieffi}, A., \& {Straniero}, O. 1993, \apj, 414, 580

\bibitem[{Searle \& Zinn(1978)}]{SearleZinn1978}
Searle, L. \& Zinn, R. 1978, The Astrophysical Journal, 225, 357

\bibitem[{{Stanway} \& {Eldridge}(2018)}]{StanwayEldridge2018}
{Stanway}, E.~R. \& {Eldridge}, J.~J. 2018, MNRAS, 479, 75

\bibitem[{Tegmark {et~al.}(1997)Tegmark, Silk, Rees, Blanchard, Abel, \&
  Palla}]{Tegmark1997}
Tegmark, M., Silk, J., Rees, M.~J., {et~al.} 1997, The Astrophysical Journal,
  474, 1

\bibitem[{Tomasetti {et~al.}(2025)Tomasetti, Moresco, Lardo, Courbin, Jimenez,
  Verde, Millon, \& Cimatti}]{Tomasetti2025}
Tomasetti, E., Moresco, M., Lardo, C., {et~al.} 2025, Astronomy \&
  Astrophysics, 699, A240

\bibitem[{{Tomasetti} {et~al.}(2026){Tomasetti}, {Moresco}, {Lardo}, {Jimenez},
  {Verde}, {Courbin}, {Millon}, \& {Cimatti}}]{Tomasetti2026}
{Tomasetti}, E., {Moresco}, M., {Lardo}, C., {et~al.} 2026, A\&A, submitted

\bibitem[{{Trenti} {et~al.}(2015){Trenti}, {Padoan}, \&
  {Jimenez}}]{Trenti2015GCagesLCDM}
{Trenti}, M., {Padoan}, P., \& {Jimenez}, R. 2015, The Astrophysical Journal
  Letters, 808, L35

\bibitem[{{Valcin} {et~al.}(2025){Valcin}, {Jimenez}, {Seljak}, \&
  {Verde}}]{valcin25}
{Valcin}, D., {Jimenez}, R., {Seljak}, U., \& {Verde}, L. 2025, \jcap, 2025,
  030

\bibitem[{{Valenzuela} {et~al.}(2025){Valenzuela}, {Forbes}, \&
  {Remus}}]{Valenzuela2025}
{Valenzuela}, L.~M., {Forbes}, D.~A., \& {Remus}, R.-S. 2025, \mnras, 537, 306

\bibitem[{Vanzella {et~al.}(2026)Vanzella, Messa, Adamo, Loiacono, Oguri,
  Sharon, Bradley, Bergamini, Meneghetti, Claeyssens, {et~al.}}]{Vanzella2026}
Vanzella, E., Messa, M., Adamo, A., {et~al.} 2026, Astronomy \& Astrophysics,
  705, A171

\bibitem[{{Wan} {et~al.}(2020){Wan}, {Lewis}, {Li}, {Simpson}, {Martell},
  {Zucker}, {Mould}, {Erkal}, {Pace}, {Mackey}, {Ji}, {Koposov}, {Kuehn},
  {Shipp}, {Balbinot}, {Bland-Hawthorn}, {Casey}, {Da Costa}, {Kafle},
  {Sharma}, \& {De Silva}}]{Wan2020}
{Wan}, Z., {Lewis}, G.~F., {Li}, T.~S., {et~al.} 2020, \nat, 583, 768

\bibitem[{{Wang} {et~al.}(2019){Wang}, {Lilly}, {Pezzulli}, \&
  {Matthee}}]{wang2019elevationsuppressionstarformation}
{Wang}, E., {Lilly}, S.~J., {Pezzulli}, G., \& {Matthee}, J. 2019, ApJ, 877,
  132

\end{thebibliography}
\end{document}